\documentclass{aa}  

\usepackage[colorlinks=true,linkcolor=blue,citecolor=blue]{hyperref}
\usepackage{graphicx}
\usepackage{txfonts}
\usepackage{caption}
\usepackage{nicematrix}
\usepackage{natbib}
\usepackage{threeparttable}
\usepackage{orcidlink} 
\newcommand{\orcid}[1]{\orcidlink{#1}}

\begin{document} 

    \titlerunning{The atmosphere of the warm Neptune-sized planet GJ 436 b with ESPRESSO}
    \authorrunning{E. Herrero-Cisneros et al. (2026)}
    
   \title{The atmosphere of the warm Neptune GJ\,436\,b probed with ESPRESSO}
   
   \author{E. Herrero-Cisneros\orcid{0000-0002-9990-6915}\inst{\ref{CAB}}\fnmsep\thanks{\href{mailto:evaherre@ucm.es}{email: \texttt{evaherre@ucm.es}}}, 
   M. R. Zapatero Osorio\orcid{0000-0001-5664-2852}\inst{\ref{CAB}}, 
   J. Sanz-Forcada\orcid{0000-0002-1600-7835}\inst{\ref{CAB}}, 
   R. Allart\orcid{0000-0002-1199-9759}\inst{\ref{OG},\ref{ITRE}}\thanks{SNSF Postdoctoral Fellow}, 
   T. Azevedo Silva\orcid{0000-0002-9379-4895}\inst{\ref{INAFA},\ref{IACE},\ref{FCUP}},
   S. Cristiani\orcid{0000-0002-2115-5234}\inst{\ref{INAFT}},
   A. R. Costa Silva\orcid{0000-0003-2245-9579}\inst{\ref{IACE},\ref{FCUP},\ref{OG}}, 
   Y. C. Damasceno\orcid{0009-0004-2297-8349}\inst{\ref{IACE},\ref{FCUP}}, 
   P. Di Marcantonio\orcid{0000-0003-3168-2289}\inst{\ref{INAFT}}, 
   P. Figueira\orcid{0000-0001-8504-283X}\inst{\ref{OG},\ref{IACE}},
   J. I. Gonz\'alez Hern\'andez\orcid{0000-0002-0264-7356}\inst{\ref{IAC},\ref{ULL}},  
   B. Lavie\orcid{0000-0001-8884-9276}\inst{\ref{OG}}, 
   M. Lendl\orcid{0000-0001-9699-1459}\inst{\ref{OG}}, 
   G. Lo Curto\orcid{0000-0002-1158-9354}\inst{\ref{ESO}}, 
   C. J. A. P. Martins\orcid{0000-0002-4886-9261}\inst{\ref{CAUP},\ref{IACE}},  
   E. Pall\'e\orcid{0000-0003-0987-1593}\inst{\ref{IAC},\ref{ULL}}, 
   F. Pepe\orcid{0000-0002-9815-773X}\inst{\ref{OG}},
   A. Psaridi\orcid{0000-0002-4797-2419}\inst{\ref{OG},\ref{ICE},\ref{IEEC}}, 
   R. Rebolo\orcid{0000-0003-3767-7085}\inst{\ref{IAC},\ref{ULL},\ref{CSIC}}, 
   J. Rodrigues\orcid{0000-0001-5164-3602}\inst{\ref{IACE},\ref{FCUP},\ref{OFXB}}, 
   N. C. Santos\orcid{0000-0003-4422-2919}\inst{\ref{IACE},\ref{FCUP}}, 
   J. V. Seidel\orcid{0000-0002-7990-9596}\inst{\ref{LL}}, 
   A. Sozzetti\orcid{0000-0002-7504-365X}\inst{\ref{INAFTorino}},
   A. Su\'arez Mascare\~no\orcid{0000-0002-3814-5323}\inst{\ref{IAC},\ref{ULL}}
   }

    \institute{Centro de Astrobiolog\'ia, CSIC-INTA, Madrid, Spain \label{CAB}
    \and D\'epartement d’Astronomie de l'Universit\'e de Gen\`eve, Chemin Pegasi 51, 1290 Versoix, Switzerland \label{OG}
    \and D\'epartement de Physique, Institut Trottier de Recherche sur les Exoplan\`etes, Universit\'e de Montr\'eal, Montr\'eal, Qu\'ebec, H3T 1J4, Canada \label{ITRE}
    \and INAF – Osservatorio Astrofisico di Arcetri, Largo Enrico Fermi 5, 50125 Firenze, Italy \label{INAFA}
    \and Instituto de Astrof\'isica e Ci\^encias do Espa\c{c}o, Universidade do Porto, CAUP, Rua das Estrelas, 4150-762 Porto, Portugal \label{IACE}
    \and Departamento de F\'isica e Astronomia, Faculdade de Ci\^encias, Universidade do Porto, Rua do Campo Alegre, 4169-007 Porto, Portugal \label{FCUP}
    \and INAF – Osservatorio Astronomico di Trieste, via G. B. Tiepolo 11, 34143, Trieste, Italy \label{INAFT}
    \and Instituto de Astrof\'isica de Canarias, C/ Vía Láctea s/n, 38200 La Laguna, Tenerife, Spain \label{IAC}
    \and Universidad de La Laguna (ULL), Departamento de Astrof\'isica, 38206 La Laguna, Tenerife, Spain \label{ULL}
    \and European Southern Observatory, Alonso de C\'ordova 3107, Vitacura, Regi\'on Metropolitana, Chile \label{ESO}
    \and Centro de Astrof\'isica da Universidade do Porto, Rua das Estrelas, 4150-762 Porto, Portugal \label{CAUP}
    \and Institute of Space Sciences (ICE, CSIC), Carrer de Can Magrans S/N, Campus UAB, Cerdanyola del Valles, 08193, Spain \label{ICE}
    \and Institut d’Estudis Espacials de Catalunya (IEEC), 08860 Castelldefels (Barcelona), Spain \label{IEEC}
    \and Consejo Superior de Investigaciones Cient\'ificas, Spain \label{CSIC}
    \and Observatoire Fran\c{c}ois-Xavier Bagnoud -- OFXB, 3961 Saint-Luc, Switzerland \label{OFXB}    
    \and Laboratoire Lagrange, Observatoire de la C\^ote d’Azur, CNRS, Universit\'e C\^ote d’Azur, Nice, France \label{LL}    
    \and INAF - Osservatorio Astrofisico di Torino, via Osservatorio 20, 10025 Pino Torinese, Italy \label{INAFTorino}
    }
    
   \date{Received -; accepted -}
 
  \abstract
   {}
   {We aim to identify the presence of atomic and molecular species in the upper atmosphere of the warm Neptune-sized transiting planet GJ\,436\,b, which has a radiative equilibrium temperature of 690 K and a mass of 25.4 $\mathrm{M_{\oplus}}$.}
   {Using the transmission spectroscopy technique, we observed two full transits of GJ\,436\,b with the ESPRESSO spectrograph, covering the wavelength range from 3800 to 7880 \r{A}. We searched for traces of atomic (H\,{\sc i}, Li\,{\sc i}, Na\,{\sc i}, Mg\,{\sc i}, V\,{\sc i}, Cr\,{\sc i}, Fe\,{\sc i}, and Fe\,{\sc ii}), along with molecular (TiO, VO) species by directly detecting planetary absorption features and by cross-correlating the planetary spectrum with theoretical spectra computed for each investigated species.}
   {Our analysis reveals no strong planetary detection for any of the species, consistent with a featureless optical spectrum. We derived upper limits by combining all ESPRESSO observations. Post-transit stellar flares were detected on both nights, primarily affecting chromospheric lines. A tentative Fe\,{\sc i} signal appears in the first transit (S/N = 3.4 $\pm$ 0.2) at a wind velocity of $\sim -18.6$ km\,s$^{-1}$, which is unexpectedly large for a cool planet. This weak signal is not present in the second transit and combined with its low significance, this suggests an origin in noise. In the less probable scenario where the feature is suppressed during the second transit by the higher stellar activity state, the T1 tentative signal peaks at 1300 K, which is above the equilibrium temperature of GJ\,436\,b. Ultimately, this result would imply a neutral iron abundance comparable to or exceeding that of the host star.}
  {}


   \maketitle
   \nolinenumbers

\section{Introduction}
\label{section:Introduction}
    
    The study of exoplanetary atmospheres provides insights into a planet's chemical composition and thermal structure, which can help constrain its formation and evolutionary history \citep[e.g.][]{Mordasini2016, Madhusudhan2017, Brewer2017}. One primary technique for characterising exoplanetary atmospheres is transmission spectroscopy, consisting of analysing the changes of the host star's spectrum as the light passes through the planet's atmosphere during a transit event. This technique has enabled the detection of signatures from a variety of molecules, as well as neutral and ionised atomic species \citep[e.g.][]{Tabernero2021, AzevedoSilva2022, Welbanks2024, Beatty2024}. Additionally, it has led to the discovery of phenomena such as temperature inversions and atmospheric winds \citep[e.g.][]{Evans2017, Hoeijmakers2018, Ehrenreich2020, Seidel2023, Wardenier2024}. The proximity of most known transiting exoplanets to their host stars exposes them to strong stellar irradiation, which not only affects their atmospheric temperature and chemistry, but also contributes to their complex evolutionary paths \citep{Brande2022}. This irradiation can cause extended atmospheres and atmospheric escape. Observations of this atmospheric escape, particularly of helium and hydrogen \citep[e.g.][]{Bourrier2018b, GarciaMunoz2019, Masson2024}, have led to the study of the interactions between a planet's atmosphere and the stellar wind, which play a significant role in shaping the exosphere and influencing the rate of atmospheric loss. Notable examples include the detection of helium in warm Neptune exoplanets like HAT-P-11\,b \citep{Allart2018, Mansfield2018} and GJ\,3470\,b \citep{Palle2020, Ninan2020}, although the latter could not be replicated by \citet{Allart2023}, indicating variability between observing nights \citep{Guilluy2024}. In addition, warm Neptunes often display flat transmission spectra in both the infrared and optical ranges, as is the case of the nearly flat optical spectrum and attenuated near-infrared molecular features in HAT-P-11 b \citep{Chachan2019}. This is commonly interpreted as the presence of high-altitude clouds or hazes that mask the absorption features. 
    
    One of the most studied exoplanetary atmospheres to date is that of the warm Neptune GJ\,436\,b \citep{Butler2004, Gillon2007}. This planet is orbiting an M3V-type star \citep{Kirkpatrick1991} with an optical brightness of $V$ =  10.6 \citep{Zacharias2013} and a near-solar metallicity, [Fe/H] = 0.10 $\pm$ 0.08 dex \citep{Rosenthal2021}. With a period of 2.64389803 $\pm$ 0.00000026 day and a radius of 3.83 $\pm$ 0.06 $\mathrm{R_{\oplus}}$ \citep{Maxted2022}, GJ\,436\,b is placed at the lower-mass edge of the hot Neptune desert \citep{Beauge2013}, making it an particularly interesting target \citep[e.g.][]{Castro-Gonzalez2024}. The detailed properties of the GJ\,436 system, as used in our study, are listed in Table~\ref{table:system}.

    \begin{table} 
    \caption{Physical and orbital parameters of the GJ\,436\,b system.}
    \setlength{\tabcolsep}{8pt}
    \begin{threeparttable}        
    \begin{tabular}{l c c}   
    \hline\hline
    \noalign{\smallskip}
    Parameter & Value & Reference \tnote{(b)} \\ 
    \hline
    \noalign{\smallskip}
         & Star & \\
    \hline 
    \noalign{\smallskip}
        $T_{\mathrm{eff}}$ (K) & 3586 $\pm$ 36 & R21 \\
        log $g$ (cgs) & 4.841 $\pm$ 0.009 & R21 \\
        $[$Fe/H$]$ (dex) & 0.10 $\pm$ 0.08 & R21 \\
        $v \sin i_{\mathrm{*}}$ ($\mathrm{m}\, \mathrm{s}^{-1}$) \tnote{(a)} & 290 $\pm$ 50 & B22 \\
        $P_{\mathrm{rot}}$ (day) & 44.09 $\pm$ 0.08 & B18 \\
        $R_{\mathrm{*}}$ ($\mathrm{R_{\odot}}$) \tnote{(a)} & 0.425 $\pm$ 0.006 & M22 \\
        $M_{\mathrm{*}}$ ($\mathrm{M_{\odot}}$) & 0.441 $\pm$ 0.009 & R21 \\
        $L_{\mathrm{*}}$ ($\mathrm{L_{\odot}}$) & 0.023 $\pm$ 0.006 & S19 \\
        $K_{\mathrm{*}}$ ($\mathrm{m}\, \mathrm{s}^{-1}$) \tnote{(a)} & 17.38 $\pm$ 0.17 & T18 \\
    \hline
    \noalign{\smallskip}
             & Planet & \\
    \hline 
    \noalign{\smallskip}
        $a$ (AU) \tnote{(a)} & 0.0286 $\pm$ 0.0002 & M22 \\
        $R_{\mathrm{p}}$ ($\mathrm{R_{\oplus}}$) \tnote{(a)} & 3.83 $\pm$ 0.06 & M22 \\
        $M_{\mathrm{p}}$ ($\mathrm{M_{\oplus}}$) & 25.4 $\pm$ 2.1 & B18 \\
        $\rho_{\mathrm{p}}$ ($\mathrm{g}\, \mathrm{cm}^{-3}$) & 2.10 $\pm$ 0.10 & This work \\
        $g_{\mathrm{p}}$ ($\mathrm{m}\, \mathrm{s}^{-2}$) & 14.11 $\pm$ 0.46 & This work \\
        $K_{\mathrm{p}}$ ($\mathrm{km}\, \mathrm{s}^{-1}$) & 100.5 $\pm$ 8.3 & This work \\
        $P$ (day) \tnote{(a)} & 2.64389803 $\pm$ 0.00000026 & L14 \\
        $T_{\mathrm{c}}$ (BJD) & 2454873.01582 $\pm$ 0.00004 & K22 \\
        $T_{\mathrm{14}}$ (h) & 1.009 $\pm$ 0.034 & B15 \\
        $i$ (deg) \tnote{(a)} & 86.8 $\pm$ 0.1 & B18 \\
        $e$ \tnote{(a)} & 0.152 $\pm$ 0.009 & T18 \\
        $\omega$ (deg) \tnote{(a)} & 325.8 $\pm$ 5.6 & T18 \\
        $b$ & 0.806 $\pm$ 0.031 & B15 \\
        $T_{\mathrm{eq}}$ (K) & 690 $\pm$ 10 & T16 \\
        \hline \\        
    \end{tabular}
    
        $^{(a)}$ Parameters used in \citet{Bourrier2022}. \\
        $^{(b)}$ R21: \citet{Rosenthal2021}; B22: \citet{Bourrier2022}; M22: \citet{Maxted2022}; S19: \citet{Stassun2019}; T18: \citet{Trifonov2018}; B18: \citet{Bourrier2018}; L14: \citet{Lanotte2014}; K22: \citet{Kokori2022}; B15: \citet{Baluev2015}; T16: \citet{Turner2016}.
    \end{threeparttable}
    \label{table:system}
    \end{table}

    The close orbit of GJ\,436\,b around its host star significantly impacts its atmospheric dynamics, primarily due to the high-energy radiation from GJ\,436, which is a driving force in the atmospheric evolution of this warm Neptune \citep{Sanz-Forcada2011}. Additionally, GJ\,436’s slightly eccentric orbit raises important questions about its dynamical history and places it in a crucial position for studying atmospheric evaporation. Notable discoveries have been made using the Space Telescope Imaging Spectrograph (STIS) on board the \textit{Hubble} Space Telescope (HST). Research by \citet{Kulow2014}, \citet{Ehrenreich2015}, and \citet{Lavie2017} revealed a deep Ly-$\alpha$ signature, indicating that GJ\,436\,b is enshrouded in a massive coma of neutral hydrogen, large enough to eclipse the star several hours before the planet's optical transit. This hydrogen cloud also forms a long cometary tail, observable for hours post-transit. Further studies by \citet{Bourrier2015, Bourrier2016} used the Ly-$\alpha$ feature to analyse interactions between the stellar wind and the planet's extended atmosphere. The persistence of this giant hydrogen exosphere has been confirmed through observations using the HST-Cosmic Origins Spectrograph (COS), as reported by \citet{dosSantos2019}, suggesting a consistent atmospheric loss over several years. Despite the observed strong Ly-$\alpha$ absorption, detecting H$\alpha$ absorption in the atmosphere of GJ\,436\,b has proven elusive \citep{Cauley2017}. Observing a planet in H$\alpha$ is more challenging than in Ly-$\alpha$, as H$\alpha$ is involved in many stellar phenomena, it resides in a spectral region with multiple stellar absorption lines, and it forms at a different temperature than Ly-$\alpha$ \citep{Jensen2012, Christie2013}.

    Furthermore, the analysis of one transit observed with the CARMENES spectrograph did not detect the He\,{\sc i} triplet at 10830 \r{A}, another potential indicator of atmospheric escape \citep{Nortmann2018}. This result is consistent with \citet{Guilluy2024}, who also did not report any evidence of the He\,{\sc i} triplet in six transits observed with the GIANO-B spectrograph. The extreme ultraviolet (EUV) flux received by the planet may not be strong enough to allow for a positive detection of the triggered He\,{\sc i} lines \citep{Oklopcic2019, Sanz-Forcada2025}.

    \citet{Knutson2014} utilised HST-Wide Field Camera 3 (WFC3) observations to measure the planetary transmission spectrum. They find a flat infrared-optical transmission spectrum, ruling out a cloud-free, hydrogen-dominated atmosphere. These results align with observations made using the HST-Near Infrared Camera and Multi-Object Spectrometer (NICMOS), which also could not find significant molecular absorption features in GJ\,436\,b's atmosphere \citep{Pont2009, Gibson2011}. \citet{Lavie2017} observed a possible absorption signal in the Si\,{\sc iii} stellar line at 1206.5 \r{A} within the HST-STIS data, which could indicate a planetary origin. If verified, this signal could imply that silicon atoms are being dragged from the lower atmosphere by hydrogen atoms, hinting at atmospheric mixing and the presence of clouds in the lower atmosphere of GJ\,436\,b \citep{Visscher2010}. However, contrasting findings were reported by \citet{Loyd2017} and \citet{dosSantos2019}, who did not detect in-transit absorption signals in the C\,{\sc ii} lines at 1334.532 and 1335.708 \r{A} and in the Si\,{\sc iii} line at 1206.5 \r{A} using HST-COS observations during a planetary transit. Several primary transits observed with the \textit{Spitzer} Space Telescope suggested that the atmosphere of this planet is mostly composed of H\textsubscript{2} and CH\textsubscript{4} \citep{Beaulieu2011}. \citet{Kesseli2020} reported a non-detection of FeH in the planet's atmosphere using high spectral resolution, near-infrared data during transit. Nevertheless, a recent analysis of one transit observed with CRIRES+ showed no transmission signals \citep{Grasser2024}.

    GJ\,436\,b's dayside has also been studied using \textit{Spitzer} secondary eclipse observations, showing a high CO abundance and traces of H\textsubscript{2}O and CO\textsubscript{2}, yet with a marked deficiency in CH\textsubscript{4} compared to what is expected from thermochemical equilibrium models for a hydrogen-dominated atmosphere \citep{Stevenson2010}. \citet{Madhusudhan2011} reanalysed the secondary eclipse data, confirming the presence of an atmosphere rich in CO and CO\textsubscript{2}, but poor in CH\textsubscript{4}. This atmospheric composition suggests a high metallicity, and possibly chemical disequilibrium processes like vertical mixing or methane polymerization \citep{Stevenson2010, Moses2013, Guzman-Mesa2022}. However, there is an absence of alkali absorption features in the low-resolution, optical transmission spectrum observed in HST-STIS transit observations \citep{Lothringer2018}, along with a non-detection of metallic ions in the HST-COS data \citep{Loyd2017, dosSantos2019}.

    Despite the large number of papers studying GJ\,436\,b’s atmosphere, there are no previous high spectral resolution observations at optical wavelengths that would allow us to identify individual spectral lines. Here, we present the first high-resolution transmission spectrum of GJ\,436\,b at visible wavelengths, acquired with the Echelle Spectrograph for Rocky Exoplanet and Stable Spectroscopic Observation (ESPRESSO; \citealt{Pepe2010, Pepe2021}). Our primary objective is to investigate the existence of atomic and molecular species with spectral features within the optical wavelength range, aiming to enhance our understanding of the temperature, composition and structure of this warm Neptune's atmosphere. This paper is organised as follows: we describe the observations in Sect.~\ref{section:Observations}. In Sect.~\ref{section:Data analysis}, we detail the data reduction and the extraction of the planetary transmission spectra. In Sect.~\ref{section:Planetary transmission spectrum}, we present the analysis of the planetary spectra and the main results are discussed in Sect.~\ref{section:Results and discussion}. The paper is concluded with a brief summary in Sect.~\ref{section:Conclusions}.

\section{Observations}
\label{section:Observations}

\subsection{Spectroscopy}
\label{section:Spectroscopy}

\begin{figure}
    \centering
    \includegraphics[width=0.75\hsize]{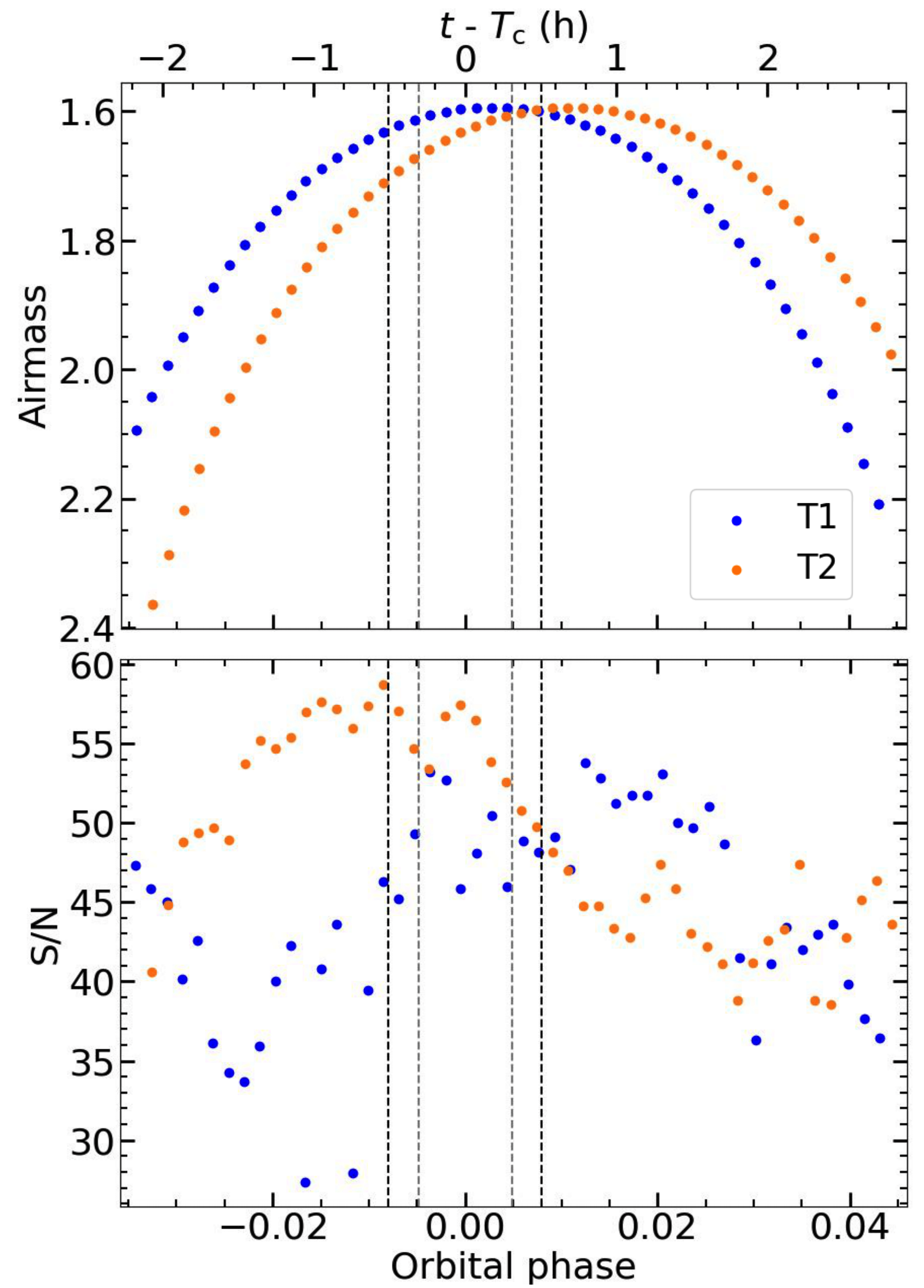}
    \caption{Variations in the airmass (\textit{top}) and S/N around 5500 \r{A} (\textit{bottom}) during T1 (blue) and T2 (orange). The vertical dashed lines indicate the orbital phases at the four contacts of the transit.}
    \label{fig:airmass_and_SN}
    \end{figure}

    Two primary planetary transits of GJ\,436\,b were observed with ESPRESSO on 2019 February 27 and April 29. The two transit dates are referred to as T1 and T2, respectively, throughout this paper. ESPRESSO \citep{Pepe2021} is an echelle fibre-fed spectrograph installed at the Very Large Telescope (VLT) of the European Southern Observatory (ESO) in Cerro Paranal, Chile. It covers a spectral range of 3771--7898 \r{A} and is particularly well-suited for high-precision transit observations. Its extreme instrumental stability and large collecting area provided by each 8.2 m radius Unit Telescope (UT) make it an ideal instrument for capturing high signal-to-noise ratio (S/N) planetary spectra. Our observations were obtained as part of the ESPRESSO Guaranteed Time Observations (program 1102.C-0744, PI: F. Pepe). The transit observations were conducted with UT3 (Melipal or the Southern Cross) using the HR mode of the ESPRESSO detector, which provides a median resolving power of $R \approx 138,000$, and the 2\,$\times$\,1 binning, suitable when the detector RON is not negligible for the error budget. Fibre A was used to observe GJ\,436 while fibre B monitored the sky contribution. A total of 49 exposures covering similar planetary orbital phases were obtained per transit event, with an individual exposure time of 300 s. Table~\ref{table:observations} provides the observing log, including the starting and ending Universal Time (UT) of the observations, the mean S/N of the data, the number of in- and out-of-transit spectra, and the planetary phases covered by the observations (with the null phase corresponding to the mid-transit times). Raw data were reduced using version 3.0.0 of the Data Reduction Software (DRS) pipeline \footnote{Available at \url{http://www.eso.org/sci/software/pipelines/index.html}} \citep{Pepe2021}. For our analysis, we employed the sky-subtracted and order-merged spectra given by the DRS, which are corrected for the Solar System barycentric velocities. Figure~\ref{fig:airmass_and_SN} shows the variations in air mass and S/N computed at 550 nm for each night.
    
    \begin{table} 
    \caption{Summary of the observations.}
    \setlength{\tabcolsep}{2.5pt}
    \small
    \begin{threeparttable}
          
    \begin{tabular}{c c c c c c c}   
    \hline\hline
    \noalign{\smallskip}
    Observing night & UT interval & $t_{\mathrm{exp}}$ & $N_{\mathrm{out}}^a$ & $N_{\mathrm{in}}^b$ & S/N$^b$ & $\Phi$ \\ 
                              &  (hh:mm)   & (s)                & & & & \\
    \hline                    
    \noalign{\smallskip}
       2019 Feb 27 (T1) & 03:32--08:27 & 300 & 39 & 10 & $\sim$44 & $-$0.034--0.043 \\     
       2019 Apr 29 (T2) & 23:08--04:01 &  300 & 39 & 10 & $\sim$52 & $-$0.033--0.044 \\ 
    \hline   
    \noalign{\smallskip}
    \noalign{\smallskip}
    \end{tabular}
    
    $^{a}$ Number of out-of-transit spectra. The number of before- and after-transit spectra are 17 (T1), 16 (T2) and 22 (T1),  23 (T2), respectively. \\
    $^{b}$ Number of in-transit spectra from first to fourth transit contacts. The number of spectra between second and third transit contacts is 6.\\
    $^{c}$ Averaged S/N measured at $\sim$5500 \r{A}.
    \end{threeparttable}
    \label{table:observations}
    \end{table}

    These observations were analysed in \citet{Bourrier2022}, which focused on the analysis of the Rossiter-McLaughlin (RM) effect to study the planetary architecture of the system, confirming the polar orbit of GJ\,436\,b. We adopt in this work the same ephemeris and planetary parameters as \citet[see our Table~\ref{table:system}]{Bourrier2022}.

    The sky emission correction in the spectral datasets is effectively handled by the ESPRESSO reduction pipeline, which performs sky subtraction for each spectrum using data from the simultaneous fibre B observations. However, the ESPRESSO DRS does not correct for telluric absorption. The Earth's atmosphere absorbs at specific optical wavelengths mostly due to the presence of $\mathrm{O}_2$, $\mathrm{H}_2\mathrm{O}$, and $\mathrm{O}_3$. In line with the methodology adopted in other ESPRESSO studies \citep[e.g.][]{Allart2020, Tabernero2021, Casasayas-Barris2022, AzevedoSilva2022, Damasceno2024, CostaSilva2024}, we corrected for this telluric contamination in each spectrum across both nights using the software \texttt{molecfit} \footnote{Available at \url{https://www.eso.org/sci/software/pipelines/skytools/molecfit}} \citep{Smette2015, Kausch2015} within the \texttt{EsoReflex} environment \citep{Freudling2013}. With \texttt{molecfit}, the barycentric Earth radial velocity (BERV) is accounted for each exposure, so the correction is performed in the Earth rest frame.

    The air wavelength regions of 6277 -- 6313 \r{A}, 6918 -- 7050 \r{A}, 7294 -- 7388 \r{A} and 7682 -- 7700 \r{A} were used for fitting the terrestrial features. These intervals were chosen as they harbour numerous, relatively weak telluric lines, while avoiding strong absorption features that could hinder the correction process. Although \texttt{molecfit} was quite effective in adjusting for microtelluric absorption \citep{Cunha2014}, it did result in significant residuals in areas of intense absorption bands, where telluric features are saturated. The telluric residuals in the intervals 6870 -- 6940 \r{A}, 7170 -- 7310 \r{A} and 7590 -- 7700 \r{A}, very much affected by strong telluric absorption, were therefore masked out in subsequent processing steps.

\subsection{Photometry}
\label{section:Photometry}

    \begin{figure}
    \centering
    \includegraphics[width=0.9\hsize]{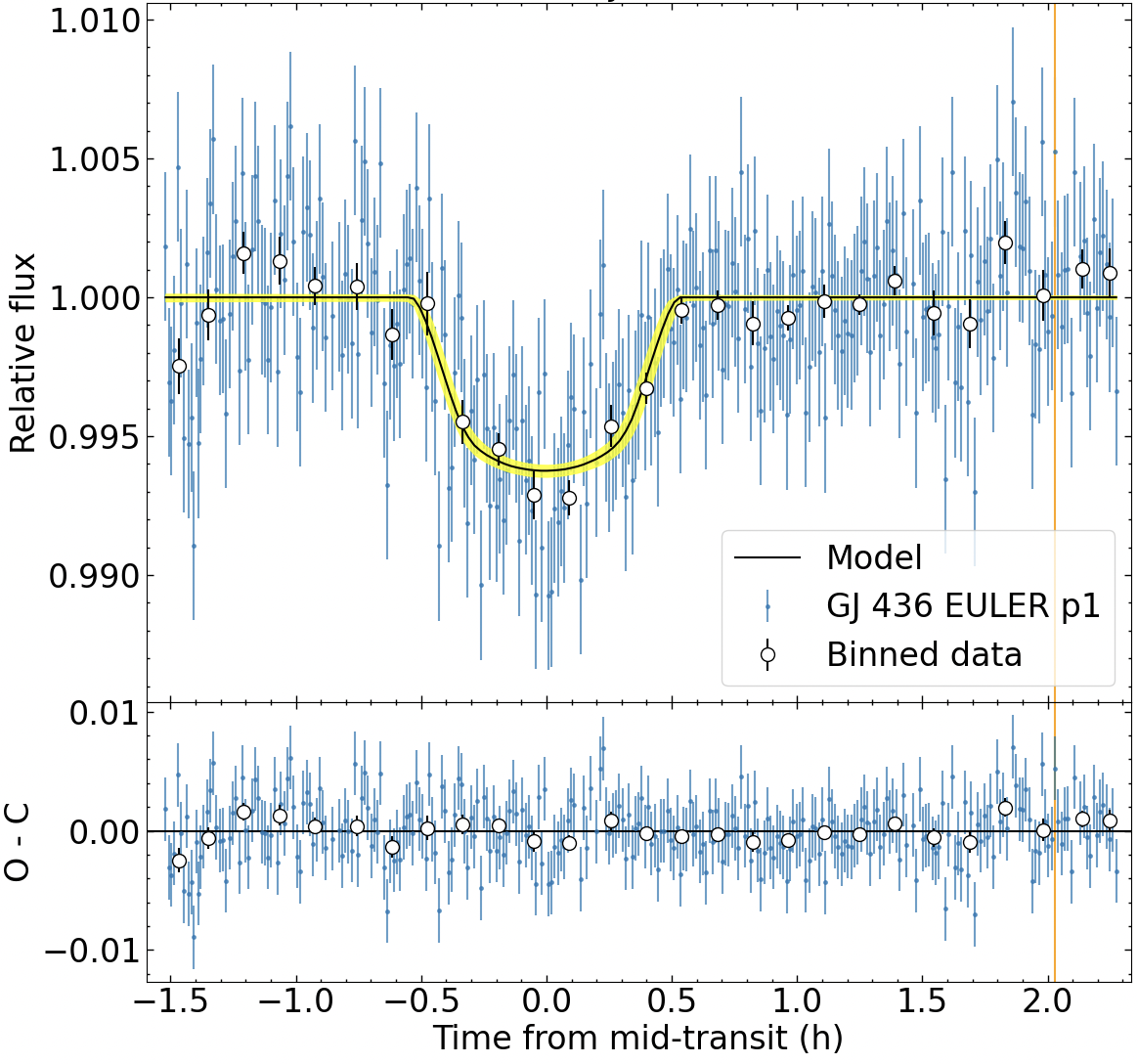}
    \caption{\textit{Top}: Euler photometry (blue dots) taken with the Gunn $r$ filter on the night of 2019 Feb 27. It is phase folded using GJ\,436\,b orbital period. The best fit model to the planetary transit and its 1$\sigma$ uncertainty are depicted by the black line and the yellow area. Binned photometry (every 11 data points) is illustrated by the white dots. The starting time of the stellar flare seen in the ESPRESSO spectra is marked by the vertical orange line. \textit{Bottom}: Photometric residuals after removing the model from the observations.}
    \label{fig:eulercam}
    \end{figure}
    
    Simultaneously with the ESPRESSO observations on the night of 2019 February 27, we also observed GJ\,436 using the EulerCAM instrument \citep{Lendl2012} of the Swiss 1.2-m Euler telescope located at ESO's La Silla Observatory in Chile. The 4k\,$\times$\,4k CCD detector of EulerCAM provides a field of view of 15\,$\times$\,15 arcmin and a plate scale of 0\farcs23 on the sky. Observations were conducted using the Gunn $r$ filter (6000--6900 \AA). Euler data were employed to confirm the times of the planetary mid-transit and first and fourth contacts, which is valuable information for identifying the ESPRESSO in- and out-of-transit spectra. Raw data were reduced following standard procedures for bias subtraction and flat-fielding. Differential photometry was obtained by comparing the flux of GJ\,436 to other point-like sources in the field. The root mean square ($rms$) of the differential photometry is 2.84 mmag (excluding the transit feature) and the typical size of the error bar per photometric measurement is 1.5 mmag. The planetary transit was successfully detected by Euler as shown in Fig.~\ref{fig:eulercam}. 
    
    We fit the Euler light curve using the \texttt{Juliet} code \citep{Espinoza19}, which employs \texttt{batman} \citep{Kreidberg15} for transit modelling. The distributions of the priors on the planetary ephemeris were normal centred on the values provided by \citet{Bourrier2022}, with standard deviations equal to the uncertainties quoted in the same work. The Euler data were flattened by fitting a straight line to the out-of-transit photometry. The fit was done together with the analysis of the planetary transit. \texttt{Juliet} was built on various nested sampling algorithms to approach the Bayesian statistics of the comparison of data to models. These algorithms sample a number of live points from the priors. In our study of the Euler data, we set the number of live points to 1000. The final fit is shown in Fig.~\ref{fig:eulercam}. We measured the mid-transit time and total duration of the transit at $T_{\mathrm{c}} = 2458542.74451^{+0.00088}_{-0.00082}$ (BJD) and $T_{\rm 14} = 1.15 \pm 0.16$ h, respectively. The difference between the Euler $T_{\mathrm{c}}$ and the value predicted by the ephemeris of Table~\ref{table:system} is 2.6 min. However, the ingress and egress times inferred from the Euler $T_{\mathrm{c}} - T_{\mathrm{14}}/2$ and $T_{\mathrm{c}} + T_{\mathrm{14}}/2$  values match within the uncertainties those derived from the \citet{Bourrier2022} ephemeris, which were used during the spectroscopic analysis.

\section{Data analysis}
\label{section:Data analysis}

\subsection{Stellar activity}
\label{section:Stellar activity}

    GJ\,436 shows periodic changes over both intermediate- and long-term scales. Its rotational period is approximately 44 days \citep{SuarezMascareno2015, Bourrier2018, DiezAlonso2019}, and it exhibits an activity cycle lasting around 7.4 years \citep{Lothringer2018, Loyd2023, Kumar2023}, as indicated by long-term photometric observations and chromospheric activity indicators. GJ\,436 also exhibits notable short-term stellar activity \citep[e.g.][]{Lothringer2018, DiezAlonso2019, Loyd2023} that we could trace during the observations using the indices provided by the ESPRESSO pipeline. These key activity tracers include the H$\alpha$ line at 6562.802 \r{A} \citep[e.g.][]{Montes1997, Robertson2016, Bourrier2018}, the Na\,{\sc i} D doublet at 5889.950 and 5895.924 \r{A} \citep[e.g.][]{Montes1997, GomesdaSilva2012, Robertson2016, Martin2017, Hintz2019}, and the Ca\,{\sc ii} H \& K lines at 3933.66 and 3968.47 \r{A} \citep[e.g.][]{Montes1996, Robertson2016, Martin2017, Hintz2019}, given in the form of the log $R'_{\mathrm{HK}}$ calibrations by \citet{Noyes1984}, a standard method for measuring chromospheric calcium emission.

\begin{figure}
    \centering
    \includegraphics[width=\hsize]{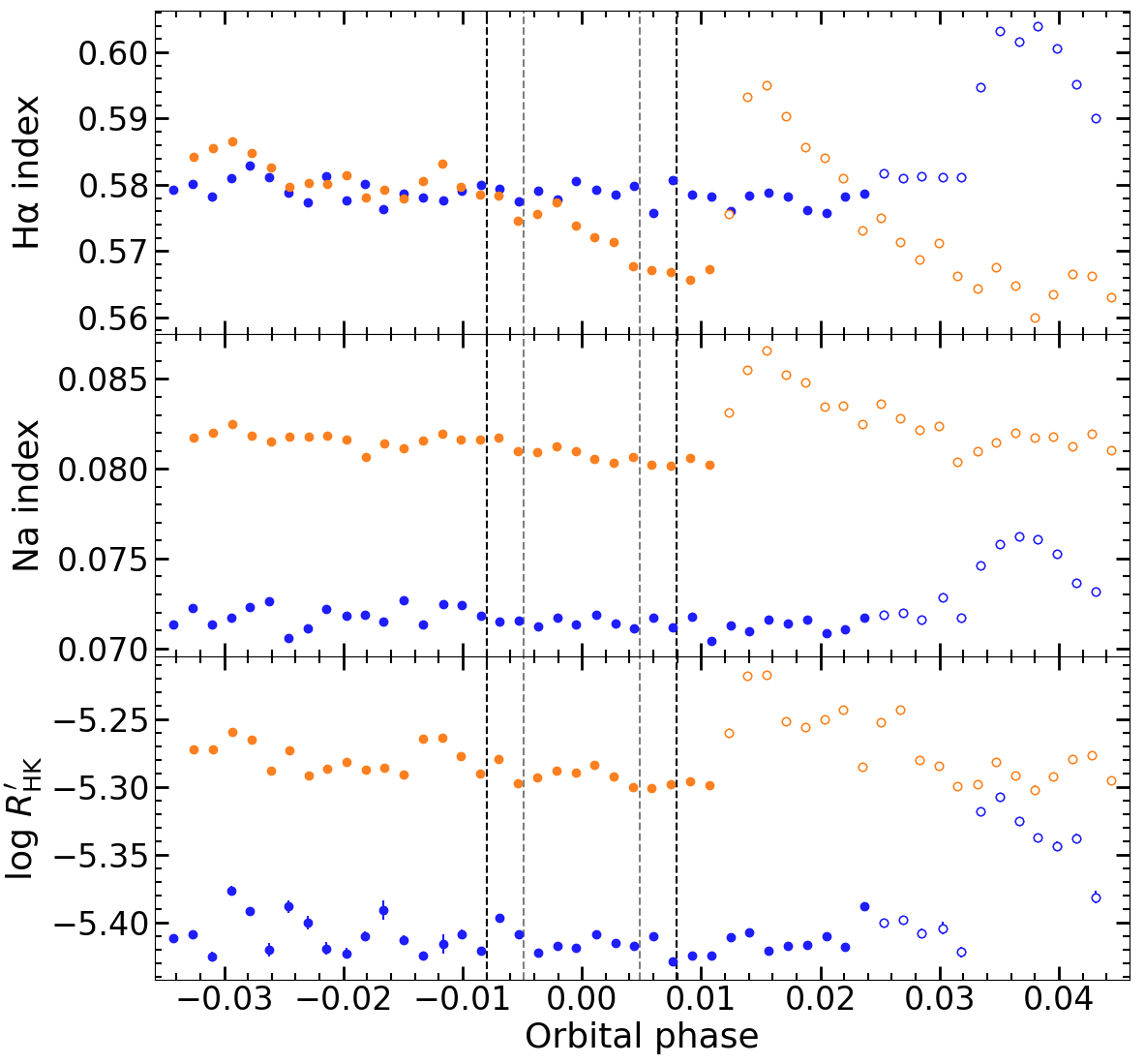}
    \caption{Variations among several stellar activity indices during the observations: H$\alpha$ (\textit{top}), Na\,{\sc i} D (\textit{middle}), and log $R'_{\mathrm{HK}}$ (\textit{bottom}). Data from T1 and T2 are shown in blue and orange, respectively. Spectra affected by stellar flares and excluded from the calculation of each night's master stellar spectrum are indicated by empty circles. Error bars are shown, but are comparable in size to the symbols. Vertical lines mark the orbital phases at the four contacts of the transit.}
    \label{fig:indices}
    \end{figure}

    Figure~\ref{fig:indices} illustrates the evolution of these stellar activity indices during ESPRESSO observations. GJ\,436 exhibited higher baseline activity levels in Na\,{\sc i} and Ca\,{\sc ii} during T2. We calculated that the mean baseline fluxes corresponding to these activity indices\footnote{Note that the H$\alpha$ and Na\,{\sc i} indices scale linearly with the stellar flux, while the Ca\,{\sc ii} indices are on the logarithmic scale.} increased by 14\% and 34\% for Na\,{\sc i} D and Ca\,{\sc ii} H \& K, respectively, from T1 to T2. 
    
    Interestingly, the spectroscopic activity indices also show distinct, brief flares that occurred shortly after the planetary transit on both observing nights. The flare observed in T1 started at approximately 07:55 UT ($\Phi$ = 0.032) and was not fully covered by the ESPRESSO observations, lasting at least 0.7 h. In T2, the flare was detected at around 02:01 UT ($\Phi$ = 0.010) and lasted for approximately 1.3 h. During the flare in T1, the stellar flux in H$\alpha$, Na\,{\sc i} D, and Ca\,{\sc ii} H \& K increased by 2.5\%, 3.7\%, and 13.5\%, respectively, compared to the flux levels before the flare. Similarly, in T2, these fluxes increased by 2.4\%, 3.9\% and 9.7\%, respectively, during the flare. These values indicate that the two flares had rather low energies at optical wavelengths, with energy significantly increasing towards shorter wavelengths. Despite covering the H$\alpha$ line, the Euler photometric data did not present any evidence of the stellar flare on T1 (Fig.~\ref{fig:eulercam}), although only the beginning of the flare (rapid rising flux) could have been registered by the Euler observations. To explore whether these low-energy flares occur at specific times relative to the planetary primary transits of GJ\,436\,b, that might suggest the presence of strong star--planet interaction, we downloaded Transiting Exoplanet Survey Satellite (TESS) light curves (Sectors 22 and 49) obtained with a cadence of 2 min \citep{Ricker2015}. We used the Pre-search Data Conditioned Simple Aperture Photometry (PDCSAP) fluxes, which were corrected for instrumental effects. We phase-folded the TESS data using the planet's ephemeris to increase the S/N and searched for flare structures immediately before and after the transits. We found no flare signature exceeding 560 ppm (1$\sigma$) sharing the same periodicity as the planet. However, this result was expected because ESPRESSO data indicate that the flare contrast strongly decreases at red wavelengths (TESS filter covers the wavelength interval 0.6--1.0 $\mu$m). Additionally, the timings of the two flares observed by ESPRESSO are similar but not identical in relation to the transit events, which likely resulted in the flare energy being diluted within the TESS photometric noise. \citet{Namizaki23} discussed that the duration of stellar flares of M dwarfs depends on the indicator, with H$\alpha$ indices having longer duration than broad-band filters. Our observation of one flare event per 5 h of continuous monitoring with ESPRESSO is consistent with the daily frequency of $\approx$10 flares observed in GJ\,436 using X-rays and far-ultraviolet (FUV) HST-COS data \citep{dosSantos2019, Loyd2023}. Regarding the extraction of the planetary spectrum, we excluded the data affected by or in close proximity to these flares from the computation of the master stellar spectra, namely, the final 12 spectra of T1 and the last 21 spectra of T2.
      
\subsection{Extraction of the planetary spectrum}
\label{section:Extraction of the planetary spectrum}

    The methodology for extracting transmission planetary spectra after correcting for telluric molecular absorption is similar to that outlined by \citet{Wyttenbach2015}. This technique has been effectively employed in high-resolution transmission spectroscopic studies, such as those carried out by \citet{Yan2017}, \citet{Allart2020}, \citet{Tabernero2021}, and \citet{Seidel2022}. All telluric-corrected spectra, previously Doppler-shifted to the Earth rest frame using the BERV indicated in the ESPRESSO pipeline, were adjusted to a uniform flux level. This step is necessary to account for differential extinction caused by variations in airmass and atmospheric transparency throughout the night. The individual spectra were normalised using ninth-order polynomials, with the lowest airmass spectrum from each night serving as the reference. This polynomial degree was chosen after testing various orders, since it best minimised the standard deviation of the residuals of the fit to the observed spectra while avoiding overfitting.

    We calculated the Keplerian model for the star using the orbital parameters listed in Table~\ref{table:system}, including the period ($P$), time of conjunction ($T_{\mathrm{c}}$), eccentricity ($e$), argument of periastron ($\omega$), and radial velocity amplitude ($K$), along with the systemic velocity, $v_{\mathrm{sys}}$. The latter parameter is strongly dependent on the technique and instrument employed for precise radial velocity (RV) determinations. Therefore, we derived $v_{\mathrm{sys}}$ independently for each night because our ESPRESSO observations were not taken with the Fabry-P\'erot simultaneous calibration on fibre B, and an RV offset was expected. The $v_{\mathrm{sys}}$ was determined as the median deviation between the theoretical Keplerian RVs (without $v_{\mathrm{sys}}$) and the ESPRESSO RVs of the out-of-transit spectra. This method intentionally excluded the in-transit spectra to avoid the influence of the 1 m\,s$^{-1}$ amplitude Rossiter-McLaughlin effect on the RVs during transit. We obtained $v_{\mathrm{sys}}$ = 9.8025 $\pm$ 0.0007  $\mathrm{km}\, \mathrm{s}^{-1}$ for T1 and 9.8015 $\pm$ 0.0005 $\mathrm{km}\, \mathrm{s}^{-1}$ for T2, implying that the ESPRESSO velocity offset was $\approx$1.0 $\pm$ 0.9 m\,s$^{-1}$ in about two months (without the Fabry-P\'erot). The quoted errors correspond to the dispersion of the flattened out-of-transit RVs after the subtraction of the Keplerian signal. They are comparable to the average error bars of the individual RV measurements, indicating that the slope of the Keplerian model accurately reproduces the ESPRESSO observations and confirming the high quality of the planetary ephemeris. However, we caution that the $v_{\mathrm{sys}}$ values reported here are applicable only to our ESPRESSO observations. Deriving absolute velocities is affected by systematic errors that are beyond the scope of this paper. Large differences (up to km\,s$^{-1}$) are observed among the $v_{\mathrm{sys}}$ values reported for GJ\,436 here and in the literature: 10.0 $\pm$ 2.5 km s$^{-1}$ \citep{upgren78}, 17.1 $\pm$ 0.6 km s$^{-1}$ \citep{dawson98}, 9.64 $\pm$ 0.14 \citep{marcy89}, 9.607 $\pm$ 0.300 km s$^{-1}$ \citep{nidever02},  9.59 $\pm$ 0.0008 km s$^{-1}$ \citep{Fouque2018}, and 8.87 $\pm$ 0.16 km\,s$^{-1}$ ({\sl Gaia} Data Release 3, \citealt{GaiaCollaboration2023}).

    The flux-corrected spectra in the Earth's rest frame were shifted to the stellar rest frame by applying the BERV from the ESPRESSO pipeline, the calculated Keplerian model, and the $v_{\mathrm{sys}}$ for each night. Subsequently, each night's master stellar spectrum was created by taking the median of the out-of-transit spectra, excluding the post-transit exposures affected by flares (Sect.~\ref{section:Stellar activity}). Accordingly, 27 and 18 exposures were employed for the T1 and T2 master stellar spectrum, respectively, indicated by empty circles in Fig.~\ref{fig:indices}. This resulted in a spectrum with S/N\,$\approx$\,220 for T1 and T2. We note that the airmass of the first three out-of-transit exposures in T2 exceeds 2.2. At such a high airmass, the ESPRESSO Atmospheric Dispersion Corrector (ADC) ceases functioning, causing atmospheric correction to stop working correctly. However, we do not expect this to significantly impact the results, as the S/N is not much lower compared to the other exposures used in the calculation of the T2 master spectrum (see Fig.~\ref{fig:airmass_and_SN}). Moreover, these three spectra represent only a small fraction of the 18 out-of-transit spectra used for the T2 master spectrum. However, to confirm this, we repeated the analysis excluding these spectra, not finding significant changes in the results presented in this paper.
    
\begin{figure}
    \centering
    \includegraphics[width=0.85\hsize]{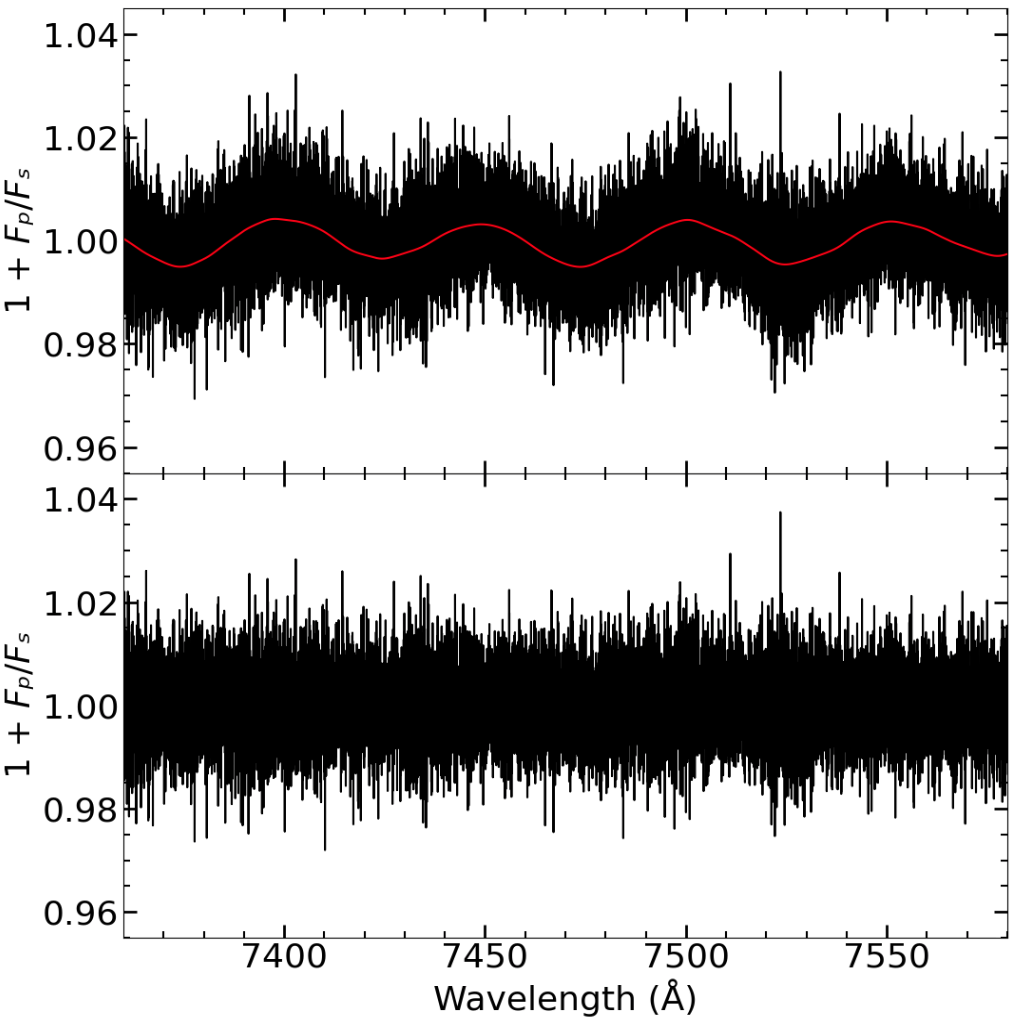}
    \caption{Example of the wiggle-correction procedure in the interval 7360 -- 7580 \r{A} in a GJ\,436\,b transmission spectrum from T1. \textit{Top}: Spectrum affected by the wiggles (black) and the fitted cubic smoothing spline function (red). \textit{Bottom}: Corrected spectrum.}
    \label{fig:correction_wiggles}
    \end{figure}

    Following this, each individual spectrum in the stellar rest frame was divided by this high S/N master spectrum to eliminate the stellar contribution from the data. At this stage of the analysis, a wiggle pattern becomes evident across the residual spectra (Fig.~\ref{fig:correction_wiggles}). This phenomenon is caused by an interference pattern of the Coudé train optics \citep[e.g.][]{Allart2020, Tabernero2021, Casasayas-Barris2021, Borsa2021, AzevedoSilva2022, Jiang2023}. The pattern exhibits variable periods throughout the ESPRESSO wavelength coverage, between $\sim$30 -- 50 \r{A}, increasing redwards, with flux amplitudes around 1\%. Due to the varying amplitude and frequency of the wiggles, cubic smoothing splines were employed to correct the data. This approach was based on the algorithm by \citet{DeBoor1978} and implemented in the \texttt{csaps} Python package \citep{Prilepin2021}. Given the high noise levels in the blue region of the transmission spectra, the wiggles were hidden in the noise. Therefore, the wiggle-correction was applied only to wavelengths greater than 4000 \r{A}. To accurately model the frequency variations in the wiggle pattern while not affecting narrow planetary features, a smoothing parameter of \texttt{smooth} = 0.0005 was set in the \texttt{csaps} function. The residual spectra were then divided by the fitted splines to produce wiggle-free residual data. Figure~\ref{fig:correction_wiggles} illustrates an example of this fitting process and the resulting wiggle-corrected spectrum. There is a second set of wiggles that have a shorter period of 1 \AA~at 6000 \AA~and a smaller amplitude of about 0.1\,\%~\citep{Casasayas-Barris2021}. Our data were not of sufficient quality to be sensitive to these low-amplitude wiggles. Therefore, we did not correct for them.

    The final step consisted of shifting the in-transit, wiggle-free residual spectra to the planetary rest frame. The planet's orbital velocity was calculated using a Keplerian model with the orbital parameters from Table~\ref{table:system}. The amplitude of the planet's orbital RV, $K_{\mathrm{p}}$, was calculated as $K_{\mathrm{p}} = K_{\mathrm{*}} \cdot M_{\mathrm{*}} \,/\, M_{\mathrm{p}}$, obtaining $K_{\mathrm{p}}$ = 100.5 $\pm$ 8.3 $\mathrm{km}\, \mathrm{s}^{-1}$, with the error derived through error propagation. We median-combined all of the in-transit spectra to improve the S/N of the data, obtaining one planetary spectrum per night. Additionally, we combined the two nights. Flux uncertainties were propagated throughout the analysis using standard analytical error propagation, based on partial derivatives and the quadratic sum of the errors.
    
\section{Planetary transmission spectrum}
\label{section:Planetary transmission spectrum}

\subsection{Atomic species}
\label{Atomic species}

\subsubsection{Narrow band}
\label{section:Narrow band}

    We searched for the H$\alpha$ line (6562.802 \r{A}), the Na\,{\sc i} D doublet (5889.950 and 5895.924 \r{A}), the Mg\,{\sc i} b triplet (5167.321, 5172.684, and 5183.604 \r{A}), and the Li\,{\sc i} doublet (6707.761 and 6707.912 \r{A}) atomic absorptions in the atmosphere of GJ\,436\,b by means of the tomography maps of the transmission spectra and the combined in-transit spectrum. These species have been identified in other planetary atmospheres, primarily in hot and ultra-hot Jupiters \citep[e.g.][]{Casasayas-Barris2018, Seidel2019, Tabernero2021, Borsa2021}. However, alkali lines have also been found in the warm Saturn WASP-127 b \citep[][]{Chen2018, Allart2020} and the hot Neptune WASP-166 b \citep[][]{Seidel2020b, Seidel2022}. We were unable to investigate the Ca\,{\sc ii} H \& K lines (3933.66 and 3968.47 \r{A}) and the K\,{\sc i} resonance doublet (7664.911 and 7698.974 \AA) due to the significant noise in the blue region ($<$4500 \r{A}) for the former, and strong O\textsubscript{2} telluric absorption in the red region of the transmission spectra for the latter. We built tomography maps in the regions around the atomic lines by vertically stacking the individual planetary spectra. These maps are shown in velocity space, where the spectral wavelength is transformed into velocities, and the theoretical position of the studied line in each case is centred at zero velocity. We did not model the centre-to-limb (CLV) variations and the RM effect in our analysis \citep[e.g.][]{Casasayas-Barris2019, Casasayas-Barris2020}, as these effects do not severely interfere with the planetary trace in this system. This is because the planetary velocities during transit (ranging from $-$17 $\mathrm{km}\, \mathrm{s}^{-1}$ to $-$8 $\mathrm{km}\, \mathrm{s}^{-1}$ in the stellar rest frame) differ significantly from the velocity span covered by the CLV and RM effects, within $\lesssim$ 1 $\mathrm{km}\, \mathrm{s}^{-1}$ around the stellar velocity, due to the low $v \sin i_{\mathrm{*}}$ ($\sim$ 0.3 $\mathrm{km}\, \mathrm{s}^{-1}$) of the star \citep{Bourrier2022}. The planetary Keplerian velocities do not intersect the zero velocity at mid-transit (left panels of Figs.~\ref{fig:Halpha}--~\ref{fig:Li}) because of the significant eccentricity of the planetary orbit ($e$ = 0.152 $\pm$ 0.009, Table~\ref{table:system}).
    
    The H$\alpha$ tomography maps depicted in the left panels of Fig.~\ref{fig:Halpha}) show notable, broad residuals along the stellar rest frame, with a width of about 50 km\,s$^{-1}$. The strong emissions observable after the transit events correlate with the stellar flares. Given the significant broadening of the residuals in the tomography maps, it was impossible to mask them without also masking any possible planetary signal. On the one hand, the H$\alpha$ tomography map of T1 does not appear to be much affected by the after-transit stellar flare. The combined in-transit planetary spectrum, shown in the top-right panel of Fig.~\ref{fig:Halpha}), is flat with an $rms$ of 3179 ppm at the position of the line. The H$\alpha$ non-detection is compatible with the findings by \citet{Cauley2017}. On the other hand, T2 appears strongly impacted by the after-transit flare, and the corresponding tomography map suggests that the ESPRESSO observations of the second transit were probably taken between two consecutive stellar flare events. The collapsed planetary spectrum of T2 (bottom-right panel of Fig.~\ref{fig:Halpha}) displays a strong, red-shifted H$\alpha$ absorption feature, whose origin is likely stellar. This will be addressed in more detail in Sect.~\ref{section:Transmission signal at H_alpha}.
            
    We combined both lines of the Na\,{\sc i} D doublet and all three lines of the Mg\,{\sc i} b triplet to strengthen the signal. Although the intensity of the individual lines of the doublet and triplet may differ, the combination has been shown to help in a few cases \citep[e.g.][]{Allart2020, Tabernero2021, Borsa2021, Seidel2023}. A stellar residual was noticeable centred around zero velocity in the stellar rest frame, more pronounced at the location of the stellar flares. We masked out the region between $\pm$ 5 km s$^{-1}$ in the stellar rest frame, effectively eliminating the stellar residual while preserving most of the potential planetary signal (left panels of Figs.~\ref{fig:Na} and ~\ref{fig:Mg}). The combined in-transit planetary spectrum did not show any planetary narrow-band signals at the Na and Mg wavelengths.

    We also searched for a potential planetary signal at the Li\,{\sc i} doublet. Because the doublet may be broadened by the planet's rotation and potential atmospheric winds, it likely remains unresolved at the resolution of ESPRESSO. Therefore, we did not combine the two line components of the  Li\,{\sc i} resonance doublet. The tomography maps showed no stellar residuals (left panels of Fig.~\ref{fig:Li}) in this case, because the host star has severely depleted Li and there is no trace of Li in the stellar spectrum. No planetary signal was detected in the combined planetary spectrum for either of the nights (Fig.~\ref{fig:Li}).

\subsubsection{Cross-correlation analysis}
\label{section:Cross-correlation analysis}

    To increase the probability of detection of atomic species in the planetary spectra, we also employed the cross-correlation function (CCF) method \citep{Snellen2010}. This technique takes advantage of many spectral features present throughout the spectra, allowing for the addition of the signal from multiple lines simultaneously by cross-correlating the observed planetary spectrum against a synthetic model containing the species of interest. The CCF technique has been widely used in recent years for studying planetary atmospheres \citep[e.g.][]{AzevedoSilva2022, Borsato2023, Pelletier2023, Prinoth2023}.

    For this purpose, we computed synthetic transmission spectra for individual species using version 2.7.7 of the high-resolution mode of the \texttt{petitRADTRANS} radiative transfer code \citep{Molliere2019}. Precomputed opacity line lists of various species are available with the code. We generated 1D plane parallel synthetic spectra for the following atomic neutral and single-ionised species: Li\,{\sc i}, Na\,{\sc i}, Mg\,{\sc i}, Mg\,{\sc ii}, Si\,{\sc i}, K\,{\sc i}, Ca\,{\sc i}, Ca\,{\sc ii}, Ti\,{\sc i}, V\,{\sc i}, Cr\,{\sc i}, Fe\,{\sc i}, and Fe\,{\sc ii}. The opacities for Na\,{\sc i} and K\,{\sc i} were taken from the Vienna Atomic Line Database (VALD; \citealt{Ryabchikova2015}). The wing shapes of the resonance line transitions of these species were computed using the theory of \citet{burrows03}. For the rest of the species\footnote{The line list database opacities available in \texttt{petitRADTRANS} can be consulted in \url{https://petitradtrans.readthedocs.io/en/latest/content/available_opacities.html}}, the opacities were calculated from the Kurucz line lists database \citep{Kurucz2018}. The range of valid temperatures for these opacities is 80 -- 4000 K.
    
    An isothermal atmospheric model was adopted, set at the planetary equilibrium temperature $T_{\mathrm{eq}}$ = 700 K \citep{Turner2016} and at a higher temperature of 1300 K (as discussed later in this section). To search for the presence of Fe\,{\sc ii}, we also generated a synthetic transmission spectrum at 3000 K, as this ion is expected to be abundant only at rather high atmospheric temperatures. Using isothermal atmospheric models is supported by the recent determination of the atmospheric pressure--temperature profile of the hot planet HD\,189733\,b by \citet{Fu2024}, indicating that an isothermal structure is a good approximation for planetary atmospheres probed with transmission spectroscopic observations. In addition, \citet{Madhusudhan2011} discussed that GJ\,436\,b does not appear to have a dayside thermal inversion. Our atmospheric model comprised cloud-free layers, a pressure range spanning from $10^{-9}$ to $10^{-3}$ bar across 150 layers, equidistant in the logarithmic space. The choice of the pressure limit of $10^{-3}$ bar is based on the capability of transmission spectroscopy to primarily probe the upper atmospheric (therefore, low pressure) layers. The lower limit of $10^{-9}$ bar approximately corresponds to the stellar radiation pressure ($P_{\mathrm{rad}}$) from GJ\,436 at a distance of the planetary orbit. $P_{\mathrm{rad}}$ was calculated using $P_{\mathrm{rad}} = L_{\mathrm{*}}^{2} / (4 \pi \, a^{2} \, c)$, with values for $L_{\mathrm{*}}$ and $a$ from Table~\ref{table:system}, while $c$ represents the speed of light in vacuum. We adopted a pressure of 10 bar for the planet's radius given in Table~\ref{table:system}, which is a required parameter by \texttt{petitRADTRANS}. This choice, representing the surface pressure at the spectral continuum, is custom in the recent literature \citep[e.g.][]{Kitzmann2023}. The volume mixing ratios of the absorbing species in the different atmospheric layers were computed under chemical equilibrium conditions, as modelled by \texttt{FASTCHEM COND} v3.1 \citep{Stock2018, Stock2022, Kitzmann2024}. We used as input the element abundances scaled to the stellar metallicity [Fe/H] = 0.1 dex \citep{Rosenthal2021}, except for lithium, for which we adopted the cosmic abundance of log\,$N$(Li) = 3.15 dex \citep{randich21} under the assumption that lithium is fully preserved in the planetary atmosphere (as opposed to the efficient lithium depletion by the host star). Lastly, we calculated the surface gravity of the planet, as it is a necessary input for the \texttt{petitRADTRANS} synthetic models, obtaining $g_{\mathrm{p}} = 14.11 \pm 0.46 \, \mathrm{m}\, \mathrm{s}^{-2}$.

    Upon examining the synthetic spectra, we find Fe\,{\sc i}, V\,{\sc i}, and Cr\,{\sc i} exhibit a high density of spectral lines in the optical at 1300 K, a temperature at which these gases are less affected by condensation. At the planet’s equilibrium temperature, only Fe\,{\sc i} displays a substantial number of absorption lines, although they are notably weaker than those present at higher temperatures (see Fig.~\ref{fig:syn_spectra_Fe_temperatures}). To correctly compare the synthetic spectra with the observational data, these synthetic models were convolved to the resolution of the ESPRESSO spectrograph (R $\sim$ 138 000) and their continua were normalised to 1 prior to the cross-correlation analysis.

    Due to the low S/N in the blue region of the transmission spectra, we excluded wavelengths below 4500 \r{A} from the analysis. The rapid decrease in S/N at these wavelengths can introduce noise-induced features in the CCFs, especially in regions densely populated with absorption lines. The CCFs were calculated in velocity space over a range of $\pm$150 km s$^{-1}$ with a step size of 0.5 km s$^{-1}$, using the \texttt{iSpec} code \citep{Blanco-Cuaresma2014, Blanco-Cuaresma2019}. This code includes error-weighted propagation based on the flux uncertainties, and follows the algorithm outlined in \citet{Pepe2002}. The CCFs were computed using the combined in-transit spectra from both nights. {The CCF analysis yielded a tentative, low confidence detection of Fe\,{\sc i} in T1 (Fig.~\ref{fig:Fe_1300_T1}; see Sect.~\ref{section:Results and discussion}). A potential velocity-coincident signal is present in the V\,{\sc i} CCF, but at lower significance than iron. We therefore did not consider it further. The resulting CCFs for V\,{\sc i} and Cr\,{\sc i} for both transits are displayed in Fig.~\ref{fig:CCF_non_detections}.

    \begin{figure*}
    \centering
    \includegraphics[width=0.9\hsize]{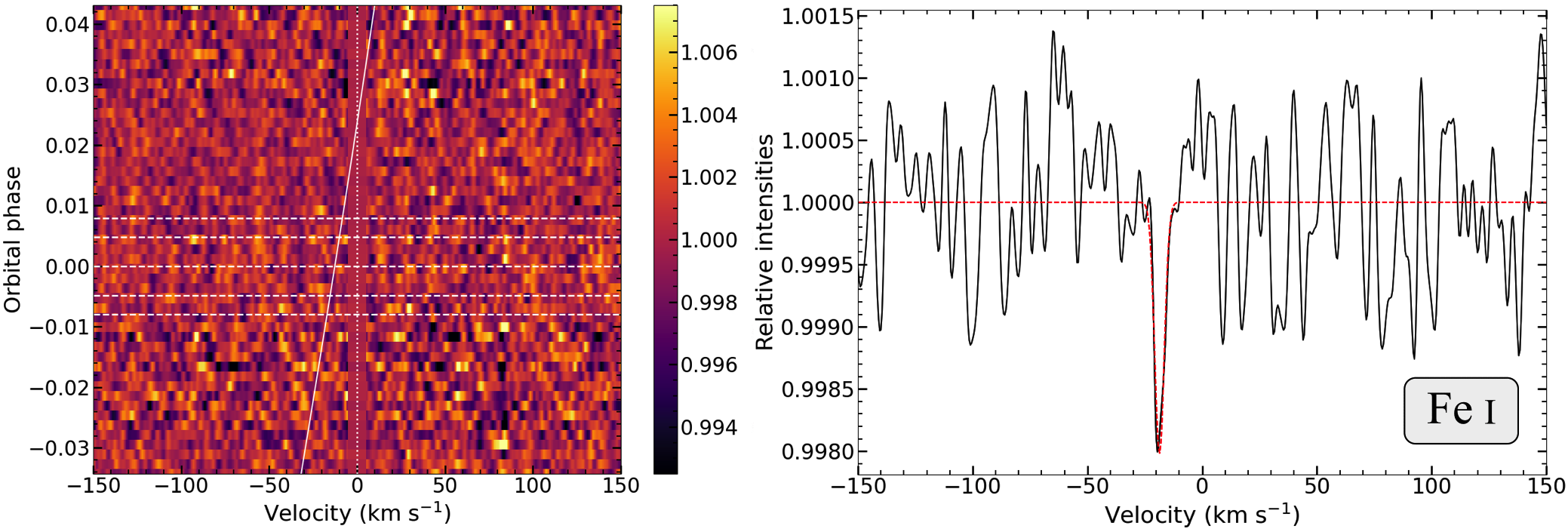}
    \caption{\textit{Left}: Tomography map of the CCFs calculated for Fe\,{\sc i} at 1300 K for T1, in the stellar rest frame. The horizontal dashed lines indicate the orbital phases at the four contacts of the transit and at midtransit, and the slanted solid line marks the planetary velocities during the observation. The dotted vertical line indicates the stellar rest frame velocity. The stellar residuals between $\pm$ 5 km s$^{-1}$ are masked out in the map. \textit{Right}: 1300 K Fe\,{\sc i} CCF of the combined in-transit planetary spectrum of T1 (solid black line), and Gaussian fit (dashed red line) of the tentative detected absorption feature. The planetary rest frame velocity is indicated by the dotted grey vertical line.}
    \label{fig:Fe_1300_T1}
    \end{figure*}

\subsection{Molecular species}
\label{section:Molecular species}

    The presence of molecular species in GJ\,436\,b's atmosphere was explored using the CCF method. Synthetic transmission spectra were generated with \texttt{petitRADTRANS} at a temperature of 1300 K for molecules showing features in the optical range at this temperature: H\textsubscript{2}O, TiO and VO. Opacity data were taken from HITEMP for H\textsubscript{2}O \citep{Rothman2010}, and from ExoMol for TiO and VO \citep{McKemmish2016, McKemmish2019}. Despite all efforts to remove telluric contribution from our ESPRESSO data (Sect.~\ref{section:Spectroscopy}), the CCFs of H\textsubscript{2}O displayed strong signatures that we ascribed to residuals of the telluric correction process. Consequently, H\textsubscript{2}O was excluded from our study. There is no evidence of the presence of TiO or VO in GJ\,436\,b's atmosphere. Figure~\ref{fig:CCF_non_detections} illustrates the resulting CCFs for TiO, and VO for both transits. Upper limits are given in Sect.~\ref{section:Upper limits of the non detections}. Our molecular non-detections are compatible with the non-detection of FeH using high-dispersion spectroscopy at near-infrared wavelengths \citep{Kesseli2020}. Non-detections could be due to different scenarios: either these gaseous species might not exist in the planetary upper atmosphere or our synthetic data were produced with poor theoretical line lists (see the next subsection); alternatively, the ESPRESSO observations might not have been sensitive enough to detect them or GJ\,436\,b has an enhanced formation of mineral haze in the form of, for instance, metal-oxide clusters (TiO$_{\rm 2}$)$_{\rm N}$, \citealt{Helling23}), which produces a general masking and thereby weakening any spectral feature \citep[e.g.][]{Fortney05}.

\section{Results and discussion}
\label{section:Results and discussion}

\subsection{H$\alpha$}
\label{section:Transmission signal at H_alpha}

    Stellar contamination is known to potentially lead to false positive detections of planetary atmospheres in transmission spectroscopy studies. For instance, the early detection of Na\,{\sc i} in HD 209458 b \citep{Charbonneau2002} was later attributed to a combination of the CLV and the RM effect \citep{Casasayas-Barris2020, Casasayas-Barris2021}. However, recent work by \citet{Carteret2024} suggests that the RM effect does not bias broadband atmospheric features, supporting the planetary origin of the Na\,{\sc i} detection by \citet{Charbonneau2002}, which would instead be probing the wings of the Na\,{\sc i} feature. Stellar variability could also be another source of contamination. To further confirm that the H$\alpha$ absorption feature observed in the residual spectrum of T2 (Fig.~\ref{fig:Halpha}) has a stellar origin, we modelled the behaviour of a Gaussian line with the instrumental width of ESPRESSO across T1 and T2 observations using the H$\alpha$ stellar activity index (Fig.~\ref{fig:indices}) provided by the ESPRESSO pipeline as a proxy. The method was described in \citet{Tabernero2022}. The results  in the form of tomography maps and collapsed planetary spectra are illustrated in Fig.~\ref{fig:Halpha_model}. The models replicate the H$\alpha$ non-detection in T1 and the existence of a red-shifted absorption feature in T2. For completeness, we applied the same modelling exercise to the first and second halves of the two transit events and compared the  results to the actual observations (Figs.~\ref{fig:Halpha_obs_and_model_T1} and~\ref{fig:Halpha_obs_and_model_T2}). This analysis again reproduced the flat planetary spectrum in the H$\alpha$ region for both halves of T1. In T2, these simulations revealed that the entire ESPRESSO observations were actually affected by stellar flare activity. The strength of the H$\alpha$ absorption feature appears more intense during the second part of the planetary transit, which is closer in time to the after-transit stellar eruption. This is likely related to an excess of H$\alpha$ absorption from the star that can occur before the flare eruption depending on the geometry, that is, where the flare is produced on the stellar surface and how it is seen with respect to the observer \citep[e.g.][]{Otsu24}. Therefore, we concluded that the observed H$\alpha$ absorption in the planetary spectrum of T2 is a product of stellar contamination.

\subsection{Upper limits on the atomic and molecular non-detections}
\label{section:Upper limits of the non detections}

    The 1$\sigma$ upper limits on the direct detections of H$\alpha$, Na\,{\sc i}, Mg\,{\sc i}, and Li\,{\sc i} lines are provided in Table~\ref{table:S1D_lines}. They were computed as the standard deviation (or dispersion) of the collapsed planetary spectrum around the position of these atomic lines, using the velocity ranges shown in Figs.~\ref{fig:Halpha}--~\ref{fig:Li}. We did not compute an upper limit on the H$\alpha$ detection of T2 because the significant, broad stellar contamination caused by the after-transit flare may have eclipsed any planetary signal. Since the quality of the two observing nights was comparable, the quoted upper limits are similar for T1 and T2. For easy comparison with the literature, we also provide in Table~\ref{table:S1D_lines} the flux upper limits in terms of planetary radius using the following expression:

    \begin{equation}
    \frac{R_{\rm eff}^2}{R_{\rm p}^2} = 1 + \frac{\sigma}{\delta}
    \end{equation}

    where $R_{\rm eff}$ is the 1$\sigma$ upper limit on the effective planetary radius for a given atomic line, $\delta$ corresponds to GJ\,436\,b transit depth, ($R_{\rm p} / R_{\rm *}$)$^2,$ and $\sigma$ stands for the calculated standard deviation. The upper limits generally tend to be more restrictive towards longer wavelengths, where the ESPRESSO spectra have higher S/N (the host star is an M dwarf, i.e. there is more flux at red than at blue wavelengths). 

    \begin{table}
    \caption[]{Upper limits (1$\sigma$) on the detectability planetary atomic lines.}
    \setlength{\tabcolsep}{4.5pt}
    \label{table:S1D_lines}
    \begin{tabular}{lccccc}
    \hline\hline
    \noalign{\smallskip}
    \multicolumn{1}{l}{Species} & 
    \multicolumn{1}{c}{$\lambda$ (\r{A})} & 
    \multicolumn{2}{c}{Flux absorption (ppm)} & 
    \multicolumn{2}{c}{$R_{\rm eff}$/$R_{\rm p}$ } \\
    \multicolumn{1}{l}{} & 
    \multicolumn{1}{c}{} & 
    \multicolumn{1}{c}{T1} & 
    \multicolumn{1}{c}{T2} & 
    \multicolumn{1}{c}{T1} & 
    \multicolumn{1}{c}{T2} \\ 
    \noalign{\smallskip}
    \hline
    \noalign{\smallskip}
    Mg\,{\sc i} b & $\sim$5175 & $\leq$ 6518  & $\leq$ 6163  & $\leq$ 1.40 & $\leq$ 1.38  \\ 
    Na\,{\sc i} D & $\sim$5890 & $\leq$ 5802  & $\leq$ 5101  & $\leq$ 1.36 & $\leq$ 1.32  \\
    H$\alpha$ & 6562.802 & $\leq$ 3179 & ---  &  $\leq$ 1.21 & ---  \\
    Li\,{\sc i} & $\sim$6707.8 & $\leq$ 4311 & $\leq$ 3834 & $\leq$ 1.28 &$\leq$ 1.25 \\ 
    \noalign{\smallskip}
    \hline
    \end{tabular}
    \smallskip  \end{table}
        
    Neutral hydrogen was identified in the atmosphere of GJ\,436\,b through Ly-$\alpha$ absorption \citep{Kulow2014, Ehrenreich2015, Bourrier2015, Bourrier2016, dosSantos2019}. This feature has also been observed in the atmospheres of similarly warm Neptunes such as GJ\,3470\,b \citep{Bourrier2018} and HAT-P-11 b \citep{Ben-Jaffel2022}, suggesting that large amounts of hydrogen could be escaping into the planetary exospheres. However, there are no reports of the H$\alpha$ detection in analogous warm Neptune exoplanets to date, the only exception being the young planet TOI-942\,c, where H$\alpha$ absorption is potentially detected at a blue-shifted velocity by \citet{Teng24}. This is likely because H$\alpha$ is significantly weaker than Ly-$\alpha$, forms at different temperatures, and because warm Neptunes have smaller sizes and reduced atmospheric scale heights, making this line harder to detect. GJ\,436\,b H$\alpha$ upper limit is a factor of $\sim$2 smaller than that of the hot Neptune TOI-2076\,b \citep{orell24} and $\sim$2 times larger than the quoted upper limit of the hot Neptune-size planet DS\,Tuc\,A\,b \citep{benatti21}. 
    
    Previous detections of Li\,{\sc i} in exoplanetary atmospheres are limited to the ultra-hot Jupiters WASP-121 b \citep{Borsa2021} and WASP-76 b \citep{Tabernero2021, Kesseli2022, Pelletier2023}, and the hot Jupiter WASP-85 A b \citep{Jiang2023}. Similarly, detections of Mg\,{\sc i} have been reported for hot and ultra-hot Jupiters \citep{Hoeijmakers2019, Prinoth2022, Tabernero2021}. To the best of our knowledge, this is the first time that these species are studied in warm Neptunes like GJ\,436\,b. There is the work by \citet{benatti21} where an upper limit on the presence of these same species is provided for the hot Neptune DS\,Tuc\,A\,b, which is $\approx$150 K hotter than GJ\,436\,b.
    
    Because of its higher chemical abundance, sodium (Na\,{\sc i} D) has been detected more frequently than the resonance lines of other alkalines in ultra-hot and hot Jupiters \citep{Seidel2019, Borsa2021, Bello-Arufe2023}. However, it is not easily detected in cooler and smaller planets. This feature has been studied in the warm Neptune HD\,106315\,c, which has $T_{\mathrm{eq}}$ = 800 K \citep{Livingston2018}. Only an upper limit of $\sim$2400 ppm was set using three planetary transit observations and data obtained with the High Accuracy Radial velocity Planet Searcher (HARPS) instrument by \citet{Zak2022}. The Na\,{\sc i} resonance doublet was detected in the warm Saturn WASP-127\,b ($T_{\mathrm{eq}} = 1400$ K, $\rho_{\mathrm{p}}$ = 0.1 $\mathrm{g}\, \mathrm{cm}^{-3}$; \citealt{Lam2017}) at a 9$\sigma$ confidence level, with an excess absorption of 0.34 $\pm$ 0.04\% \citep[][]{Allart2020}. Also, in the warm, bloated Neptune WASP-166\,b ($T_{\mathrm{eq}} = 1270$ K, $\rho_{\mathrm{p}} = 0.5, \mathrm{g,cm}^{-3}$; \citealt{Hellier2019}), Na\,{\sc i} D was detected with an absorption intensity of 4550 $\pm$ 1350 ppm. \citet{Seidel2020b, Seidel2022} also reported Na\,{\sc i} detection in WASP-166\,b at 3.4$\sigma$ confidence. Our Na\,{\sc i} D detectability level is comparable to these signals. However, since GJ\,436\,b is about twice as cool as WASP-127\,b and WASP-166\,b, the Na\,{\sc i} D feature is expected to be significantly smaller, making the detection of these individual lines particularly challenging at the current S/N level. Additional transit observations with ESPRESSO would be necessary to reach the sensitivity required to detect this narrow doublet.

    \begin{table}
    \caption[]{1$\sigma$ upper limits on the detectability of various species using the CCF method.}
    \centering
    \setlength{\tabcolsep}{5pt}
    \label{table:CCF_non_detections}
    \begin{tabular}{lcccc}
    \hline\hline
    \noalign{\smallskip}
    \multicolumn{1}{l}{Species} & 
    \multicolumn{2}{c}{Flux absorption (ppm)} & 
    \multicolumn{2}{c}{$R_{\rm eff}$/$R_{\rm p}$ } \\
    \multicolumn{1}{l}{} & 
    \multicolumn{1}{c}{T1} & 
    \multicolumn{1}{c}{T2} & 
    \multicolumn{1}{c}{T1} & 
    \multicolumn{1}{c}{T2} \\
    \noalign{\smallskip}
    \hline
    \noalign{\smallskip}
    V\,{\sc i}    & $\le$ 1728 & $\le$ 1551 & $\le$ 1.120  & $\le$ 1.108  \\
    Cr\,{\sc i}   & $\le$ 957 & $\le$ 937  & $\le$ 1.068 & $\le$ 1.067 \\
    Fe\,{\sc ii}  & $\le$ 1856  & $\le$ 1643 & $\le$ 1.128  & $\le$ 1.114  \\
    TiO & $\le$ 28  &  $\le$ 26 & $\le$ 1.002  & $\le$ 1.002 \\
    VO & $\le$ 35   & $\le$ 29  & $\le$ 1.003 & $\le$ 1.002 \\
    \noalign{\smallskip}
    \hline
    \end{tabular}
    \tablefoot{Synthetic spectra were generated at a temperature of 1300 K, except for Fe\,{\sc ii}, which was modelled at 3000 K.}
    \\
    \end{table}
    
    We also determined 1$\sigma$ upper limits on the detectability of various atomic and molecular species using the CCF technique (Sects.~\ref{section:Cross-correlation analysis} and~\ref{section:Molecular species}) and the combined planetary spectrum from both nights. These upper limits, derived from the dispersion of the CCFs, are summarised in Table~\ref{table:CCF_non_detections}. This method provides enhanced sensitivities because it takes advantage of a large number of absorption lines per species (e.g. see the synthetic spectra shown in Fig.~\ref{fig:syn_spectra_1300K}). However, it is highly sensitive to the quality of the atomic and molecular opacities used for producing the synthetic spectra. If the positions of the atomic and molecular transitions are wrongly predicted and/or the catalogue of transitions is incomplete, the CCF method will yield unreliable results. \citet{Gharib-Nezhad2021} explored the effect of absorption cross-sections calculated from different line lists in the context of ultra-hot Jupiter and M-dwarf atmospheres finding significant variations in the theoretical, high-resolution transmission and emission spectra. These authors highlighted that the most significant differences arise from the choice of the TiO line lists below 1 $\mu$m. More recently, \citet{McKemmish2024} have shown that significant improvement on the line lists of diatomic molecules like MgO, VO, and TiO is still possible. This led us to remain cautious when considering the results related to the oxides and possibly the hydrides as well, which are shown in Table~\ref{table:CCF_non_detections}. The line lists and opacities of atomic species have better constraints published in the literature \citep[e.g.][]{Ryabchikova2015, Kurucz2018}. Therefore, we considered the upper limits imposed on the atomic species to be more reliable.

\subsection{Marginal evidence for Fe\,{\sc i}?}
\label{Marginal evidence for Fe?}

    The CCF computed for Fe\,{\sc i} at the planet's equilibrium temperature ($T_{\mathrm{eq}} \approx$ 700 K) and stellar metallicity ([Fe/H] = 0.1 dex) revealed a marginal signal during the first transit. To further characterise this potential atomic signal, which would represent the first detection in a warm Neptune-like planet if confirmed, we generated additional Fe\,{\sc i} synthetic transmission spectra using isothermal atmospheric models spanning temperatures from 700 K to 2000 K in increments of 100 K, adopting the stellar metallicity. We explored this extended temperature range to account for upper atmospheric layers hotter than the equilibrium temperature due to thermal inversions likely caused by strong optical absorbers and/or stellar irradiation \citep[e.g.][]{Sheppard2017, Arcangeli2018}.

    The observed T1 spectrum of GJ\,436\,b was then cross-correlated against all newly-computed synthetic spectra. The Fe\,{\sc i} signature appears to be present at the same velocity in all new CCFs. To identify what temperature maximises the potential signal, we fitted a Gaussian function to the absorption profile in each CCF, thereby obtaining the parameters for absorption depth ($h$), full width at half maximum (FWHM), and velocity shift ($v_{\mathrm{wind}}$). We defined the S/N of the putative signal as the value of $h$ divided by its error, the latter computed as the standard deviation of the CCF between $-$150 and $+$150 km s$^{-1}$ after masking the absorption signature. We assigned an uncertainty to the derived S/N values based on the dispersion of 1000 possible S/N measurements by adopting different smaller velocity intervals (of at least 50 km s$^{-1}$) for computing the CCF standard deviations. The errors for FWHM and $v_{\mathrm{wind}}$ were determined by fitting Gaussian profiles fixing the continuum both above and below half the noise level from the average continuum level, and comparing their fitted Gaussian parameters to the central values previously determined. Figure~\ref{fig:Fe_SN} displays the final S/N of the identified potential signals as a function of the temperature of the models. The S/N exhibits a temperature-dependent trend, peaking at $\sim$1300 K. The low S/N at temperatures cooler than 1200 K is attributed to the depletion of gaseous Fe\,{\sc i} due to higher condensation at these lower temperatures. We determined S/N =  3.4 $\pm$ 0.2 for the putative Fe\,{\sc i} signal, indicating low statistical significance and insufficient evidence for a detection. Table~\ref{table:results_Fe_detection} summarises the CCF measurements for T1.

    \begin{figure}
    \centering
    \includegraphics[width=0.9\hsize]{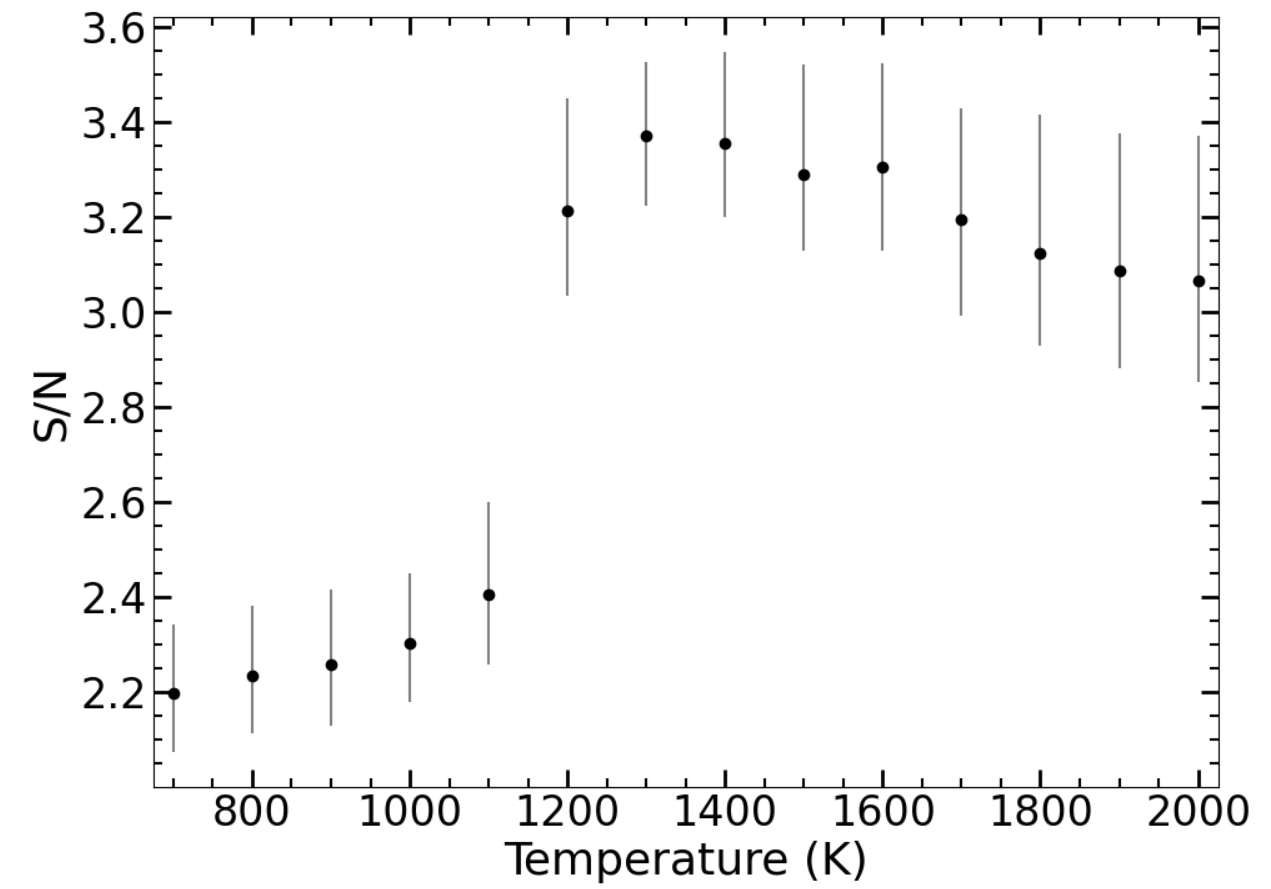}
    \caption{S/N of the Fe\,{\sc i} signal detected in T1 as a function of the temperature of the synthetic template spectra (with fixed [Fe/H] = $+0.1$ dex). The S/N of the detection peaks at $\sim$1300 K.}
    \label{fig:Fe_SN}
    \end{figure}
    
    \begin{table}
    \caption[]{CCF measurements for the tentative Fe\,{\sc i} signal.}
    \centering
    \setlength{\tabcolsep}{5pt}
    \label{table:results_Fe_detection}
    \begin{tabular}{lcc}
    \hline\hline
    \noalign{\smallskip}
    Parameter & Fe\,{\sc i}\\
    \noalign{\smallskip}
    \hline
    \noalign{\smallskip}
    $h$ T1 (ppm) & 2019 $\pm$ 598\\
    $h$ T2 (ppm) &  $\le$ 720 \\
    $R_{\rm eff}/R_{\rm p}$ T1 & 1.14 $\pm$ 0.04  \\
    $R_{\rm eff}/R_{\rm p}$ T2 &   $\le$  1.05 \\
    $v_{\mathrm{wind}}$ T1 (km s$^{-1}$) & $-$18.6 $\pm$ 0.4 \\
    FWHM T1 (km s$^{-1}$) & 5.0 $\pm$ 0.8 \\
    S/N T1  & 3.4 $\pm$ 0.2 \\
    \noalign{\smallskip}
    \hline
    \end{tabular}
    \end{table}

    To assess the reliability of the weak Fe\,{\sc i} CCF signal in the planetary rest frame, we constructed the $K_{\mathrm{p}}$--$v_{\mathrm{sys}}$ maps for T1 and T2 using the CCFs obtained as described in Sect.~\ref{section:Cross-correlation analysis} for each $K_{\mathrm{p}}$ velocity. The resulting maps are displayed in Fig.~\ref{fig:Kp_map}. In T2, the strongest signal occurs at around $K_{\mathrm{p}}$ = 0 and $v_{\mathrm{sys}}$ = 0 km\,s$^{-1}$, likely driven by increased stellar variability during the second transit. In the T1 map, the marginal signal shown in Fig.~\ref{fig:Fe_1300_T1} lies along an extended inclined structure spanning a broad range of $K_{\mathrm{p}}$ velocities, casting doubt on its planetary origin. The structure shows a slightly enhanced S/N near $K_{\mathrm{p}}$ = 110 km\,s$^{-1}$, consistent with the expected planetary value within 1--2~$\sigma$ (Table~\ref{table:system}).

    To evaluate whether the weak signal might arise from data noise or systematics, we performed an empirical Monte Carlo (EMC) or bootstrapping diagnostic, following the method described in \citet{Redfield2008}. This approach involved randomly selecting individual exposures to build an 'in-transit' sample and another 'out-of-transit' sample. From these, we extracted the transmission spectrum similarly as described in Sect.~\ref{section:Extraction of the planetary spectrum}. We then calculated the Fe\,{\sc i} CCFs following the procedure outlined in Sect.~\ref{section:Cross-correlation analysis}. Gaussian profiles were fitted at the expected absorption velocity (see Table~\ref{table:results_Fe_detection}) and the absorption depth was measured for each sample. We considered various scenarios similar to those described in \citet{Redfield2008} and commonly used in other studies \citep[e.g.][]{Wyttenbach2015, Casasayas-Barris2019, Damasceno2024}. In the first scenario (out--out), out-of-transit observations were randomly split into two equal-sized samples, one serving as the in-transit sample and the other as the out-of-transit sample. The second scenario (in--in) was constructed similarly by splitting the in-transit observations. In the in--out scenario, the in-transit and out-of-transit samples were built using in- and out-of-transit observations, respectively. Lastly, we considered an additional mixed scenario, where both in- and out-of-transit data were randomly combined to construct the in-transit and out-of-transit samples. In this last scenario, we maintained the same proportion between the number of in and out exposures of T1 (see Sect.~\ref{section:Extraction of the planetary spectrum}), that is, ten exposures were used for the in-transit sample and the remaining data were assigned to the out-of-transit sample. We note that exposures affected by the stellar flare (the final 12 spectra of the night) were excluded from the analysis in all EMC scenarios.

    We generated 20,000 random samples for each scenario. The resulting absorption depth distributions are shown in Fig.~\ref{fig:EMC}. The in--out scenario yielded a distribution centred at 2060 $\pm$ 390 ppm, clearly shifted from zero. In contrast, the in--in, out--out, and mixed scenarios produced distributions centred at 500 $\pm$ 810 ppm, $-$320 $\pm$ 570 ppm and 250 $\pm$ 630 ppm, respectively, all compatible with zero at 1 $\sigma$. The mixed scenario was used to estimate a small false-positive probability of 0.12\%~for the Fe\,{\sc i} signal. We note that the bootstrap samples are not statistically independent due to the limited number of exposures ($\sim$40). However, this exercise showed that only the in–out configuration exhibits a significant deviation from zero. This result suggests that the observed Fe\,{\sc i} signature originates from in-transit exposures rather than from a random arrangement, although it does not alone robustly confirm a planetary origin.
        
    \begin{figure}
    \centering
    \includegraphics[width=0.8\hsize]{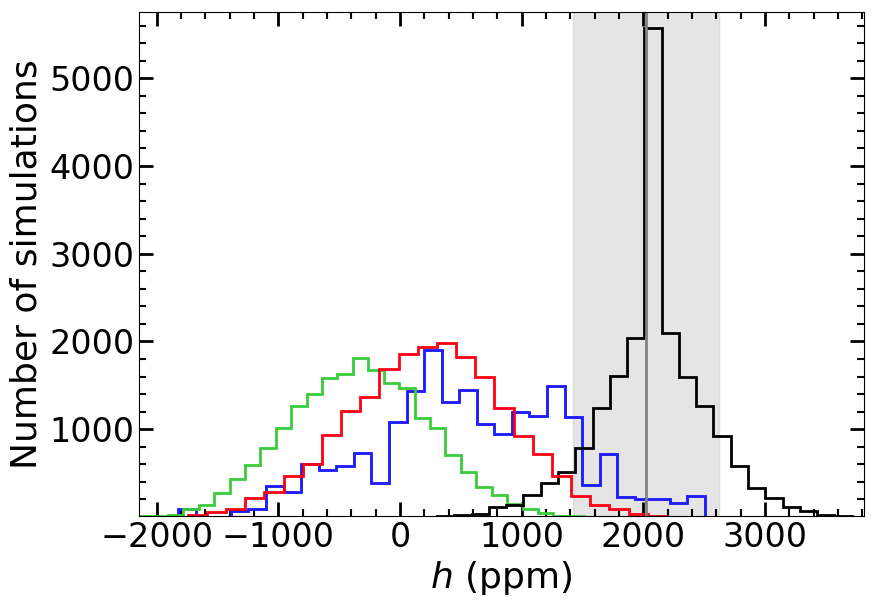}
    \caption{Distributions of the absorption depth ($h$) from the bootstrap analysis of the Fe\,{\sc i} CCFs. The in--in distribution is shown in blue, out--out in green, mixed in red, and in--out in black. The measured $h$ value from the CCF detection and its 1$\sigma$ error (see Table~\ref{table:results_Fe_detection}) are indicated by the grey vertical line and shaded region.}
    \label{fig:EMC}
    \end{figure}

\subsection{Fe\,{\sc i} versus Na\,{\sc i}}
\label{section:Fe_Na_detectability}

    The observed marginal Fe\,{\sc i} signal may appear inconsistent with the non-detection of Na\,{\sc i}, given that the Na\,{\sc i} D lines at $\sim$5890 \AA~have much higher opacity than individual Fe\,{\sc i} lines. However, the synthetic spectra computed in Sect.~\ref{section:Cross-correlation analysis} predict Na\,{\sc i} D line depths of $\sim$1200 ppm (700 K) and $\sim$2900 ppm (1300 K), still below our upper limits (Table~\ref{table:S1D_lines}; Sect.~\ref{section:Upper limits of the non detections}) by a factor of a few or more. This indicates that based on theoretical predictions the ESPRESSO data lack the sensitivity required to detect Na\,{\sc i} D in the atmosphere of GJ\,436\,b.

    To investigate whether Na\,{\sc i} and Fe\,{\sc i} could be detected via the CCF technique, we performed an injection-recovery study using synthetic planetary transmission spectra computed at 1300 K for varying abundances (Sect.~\ref{section:Cross-correlation analysis}). Neutral iron and sodium abundances were set to the stellar values and [X/H] = 0.5, 1.0, 1.5, and 2.0 dex. Wavelength-dependent Gaussian noise was added in 30 \AA~chunks, matching the noise levels in the T1 transmission spectrum. For each abundance, 10,000 noisy spectra were generated to adequately sample the noise. CCFs were then computed by cross-correlating these models with the noiseless Fe\,{\sc i} or Na\,{\sc i} spectra at 1300 K (stellar abundance), and the S/N of the feature at the planetary rest frame was measured. The resulting S/N distributions are shown in Fig.~\ref{fig:histograms_Fe_syn_detection_T1} (Fe\,{\sc i}) and Fig.~\ref{fig:histograms_Na_syn_detection_T1} (Na\,{\sc i}).

    The predicted Na\,{\sc i} S/N distributions are insensitive to abundance over [Na/H] = 0.1--2 dex. Visual inspection of the synthetic spectra revealed that only the Na\,{\sc i} D line wings vary appreciably with abundance, producing small impact on the CCFs. Consequently, no significant Na\,{\sc i} signal (S/N $>$ 3) is recovered at any sodium abundance. In contrast, the predicted Fe\,{\sc i} S/N increases steadily with iron abundance, as higher abundances produce deeper spectral features. As shown in Fig.~\ref{fig:histograms_Fe_syn_detection_T1}, high iron abundances ([Fe/H] $>$ 0.5 dex) yield predicted S/N $>$ 3. This injection-recovery analysis indicates that, under the adopted assumptions and noise conditions, Na\,{\sc i} is less detectable than Fe\,{\sc i} via the CCF technique, despite the intrinsically higher opacity of the sodium resonance lines.

    \begin{figure}
    \centering
    \includegraphics[width=0.8\hsize]{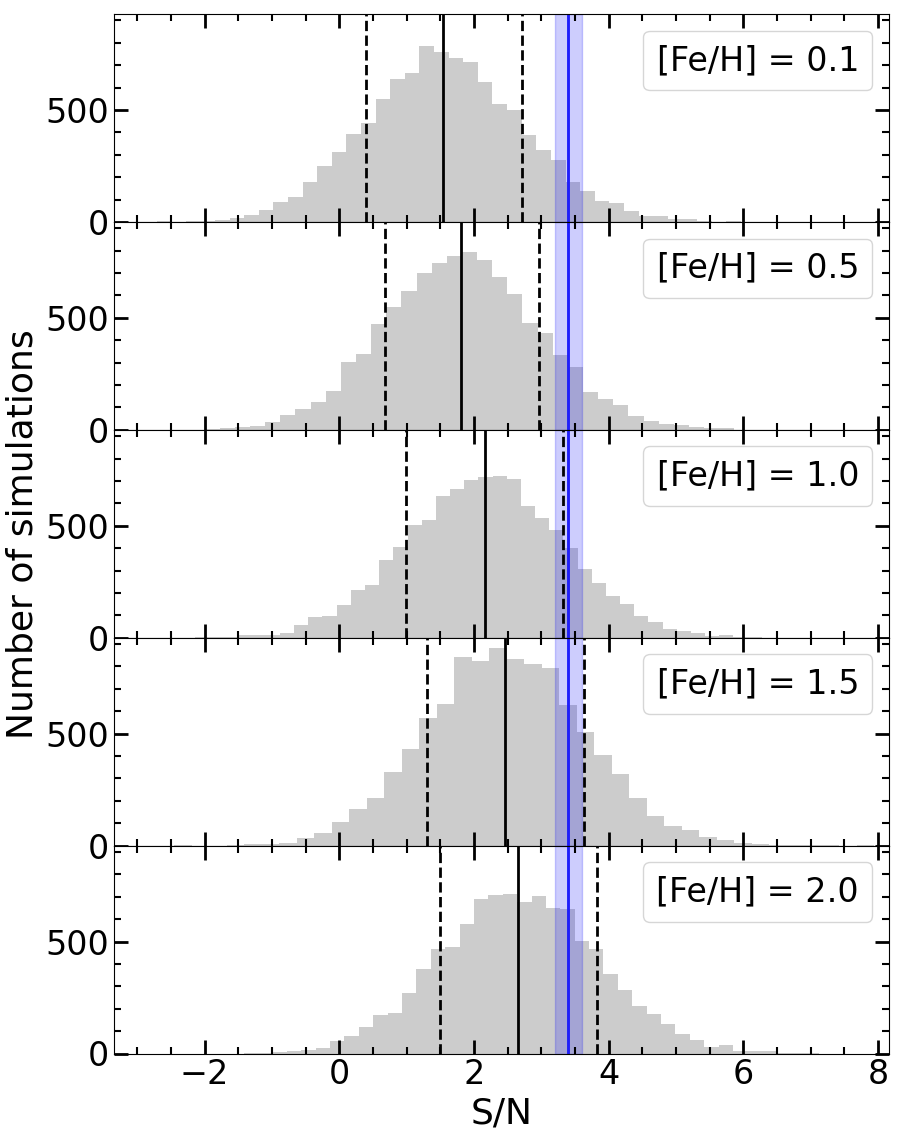}
    \caption{Distribution of the S/N of Fe\,{\sc i} signals using the CCF method and synthetic spectra computed for different metallicities (stellar metallicity is [Fe/H] = 0.1 dex). The peak of the distribution lies at the 50$^{\rm th}$ percentile (solid black line), while the 16$^{\rm th}$ and 84$^{\rm th}$ percentiles are shown by the dashed black lines. The measured S/N of the tentative Fe\,{\sc i} detection in GJ\,436\,b's atmosphere using the ESPRESSO observations (S/N = 3.4 $\pm$ 0.2) is indicated by the blue band.}
    \label{fig:histograms_Fe_syn_detection_T1}
    \end{figure}

    It is important to acknowledge the challenges in computing reliable planetary spectra for comparison with observations. In addition to the loss of the planetary continuum in high-resolution spectroscopy, an inherent consequence of the spectral extraction technique \citep[e.g.][]{Birkby2018, Seidel2020a, Maguire2023}, uncertainties also arise from shortcomings in the model spectra, such as the neglect of non-local thermodynamic equilibrium (NLTE) effects \citep[e.g.][]{GarciaMunoz2019, Lampon2020, Czesla2022} and the omission of disequilibrium chemistry and clouds in our theoretical data. The software \texttt{petitRADTRANS} allowed us to explore the impact of adding opaque clouds into GJ\,436\,b's atmosphere using the grey approximation, assuming cloud opacity to be wavelength-independent. The exact location of the cloud deck is uncertain; however, \citet{Knutson2014}, \citet{Grasser2024}, and  \citet{Finnerty2026} suggested it may reside at pressures around 10$^{-3}$ bar. We generated models with cloud decks at 10$^{-3}$, 10$^{-4}$, and 10$^{-5}$\,bar, finding that the deepest layer had no effect on the strength of the iron lines, as it likely lies beneath the atmospheric regions probed by optical observations. For decks at 10$^{-4}$ and 10$^{-5}$\,bar, predicted S/N $\sim$ 2--3 requires iron abundances of $\sim$10$\times$ and $\sim$100$\times$ solar, respectively.

    Another source of uncertainty is the planetary pressure--temperature profile. Any estimate of the neutral iron abundance is highly sensitive to the adopted vertical structure. The choice of the outer boundary (pressures below $10^{-6}$ bar) has little impact on Fe\,{\sc i}, as iron, being heavy, resides in deeper layers. In contrast, the definition of the inner layers strongly affects the simulations. We adopted $10^{-3}$ bar (Sect.~\ref{section:Cross-correlation analysis}), consistent with studies showing that ESPRESSO wavelengths probe pressures of $10^{-3} - 10^{-6}$ bar  in hot Jupiters with $T_{\rm eq} \sim 1500$ K \citep{Maguire2023}, while JWST data indicate infrared wavelengths probe $10^{-2} - 10^{-4}$ bar in similar planets \citep[e.g.][]{Fu2024}. If the deepest layers probed by ESPRESSO were at higher pressures ($10^{-2}$ bar), Fe\,{\sc i} lines would be stronger, and lower iron abundances would be needed to reach predicted S/N = 2--3. Even when including these deeper layers, our simulations indicate that an Fe-depleted planetary atmosphere ([Fe/H] $< -0.5$ dex) cannot reach a predicted S/N $>$ 3.

\subsection{Comparison with previous results}
\label{section:Comparison with previous results}

    Although the tentative planet signal in Fig.~\ref{fig:Fe_1300_T1} cannot be confirmed with the current data, we discussed below the potential implications for GJ\,436\,b's atmosphere if it were validated by future observations.

    The derived S/N of the potential planetary Fe\,{\sc i} signal from T1 is indicated in Fig.~\ref{fig:histograms_Fe_syn_detection_T1}. From this figure,  [Fe/H] $\ge$ 1 would be required to reproduce the observed tentative Fe\,{\sc i} signal, i.e. $\ge$8 times the stellar iron abundance. If the signal were confirmed, this suggests that the outer atmosphere of GJ\,436\,b could be enriched in neutral iron relative to its host star, consistent with \textit{Spitzer} findings of a dayside metallicity $\sim$10$\times$ solar \citep{Stevenson2010, Madhusudhan2011}. However, the low statistical significance of the ESPRESSO Fe\,{\sc i} signal prevents firm conclusions. In addition, the simulations assumed chemical equilibrium without photochemistry. Super stellar metal enrichment is commonly observed in giant exoplanets using transmission and emission spectroscopy (e.g. WASP-127\,b, \citealt{Kanumalla24}; HD\,189733\,b, \citealt{Fu2024}; HD\,149026\,b, \citealt{Bean2023}), with HD\,149026\,b reaching 59--276$\times$ solar metallicity (the most metal-rich planet known so far). The planet--star metallicity discrepancy may reflect formation history \citep{Booth2017,VanderMarel2021,Molliere2022} and/or hydrogen loss during atmospheric escape \citep{Ehrenreich2015, Bourrier2016, Lavie2017}, which enriches heavier species like iron over time.

    Previous Fe\,{\sc i} detections in exoplanetary atmospheres have so far been limited to hot and ultra-hot Jupiter planets \citep[e.g.][]{Ishizuka2021, Tabernero2021, Prinoth2022, AzevedoSilva2022, Bello-Arufe2022, Prinoth2023, Seidel2023}. Figure~\ref{fig:Fe_detections} summarises all reported Fe\,{\sc i} planetary detections from transmission spectroscopy observations according to the ExoAtmospheres database\footnote{\url{https://research.iac.es/proyecto/exoatmospheres/index.php}}. There are a total of 10 planets, with bulk densities spanning a factor of ten. There is no obvious trend with planetary bulk density or temperature. Neutral iron in the gas form is typically seen close to 1 R$_{\rm p}$. Two planets, KELT-9\,b and HD\,85628A\,b, have different $R_{\rm eff}$ determinations from the literature (all shown in Fig.~\ref{fig:Fe_detections}). GJ\,436\,b is included for comparison purposes; it does not deviate from the observed pattern.

    \begin{figure}
    \centering
    \includegraphics[width=\hsize]{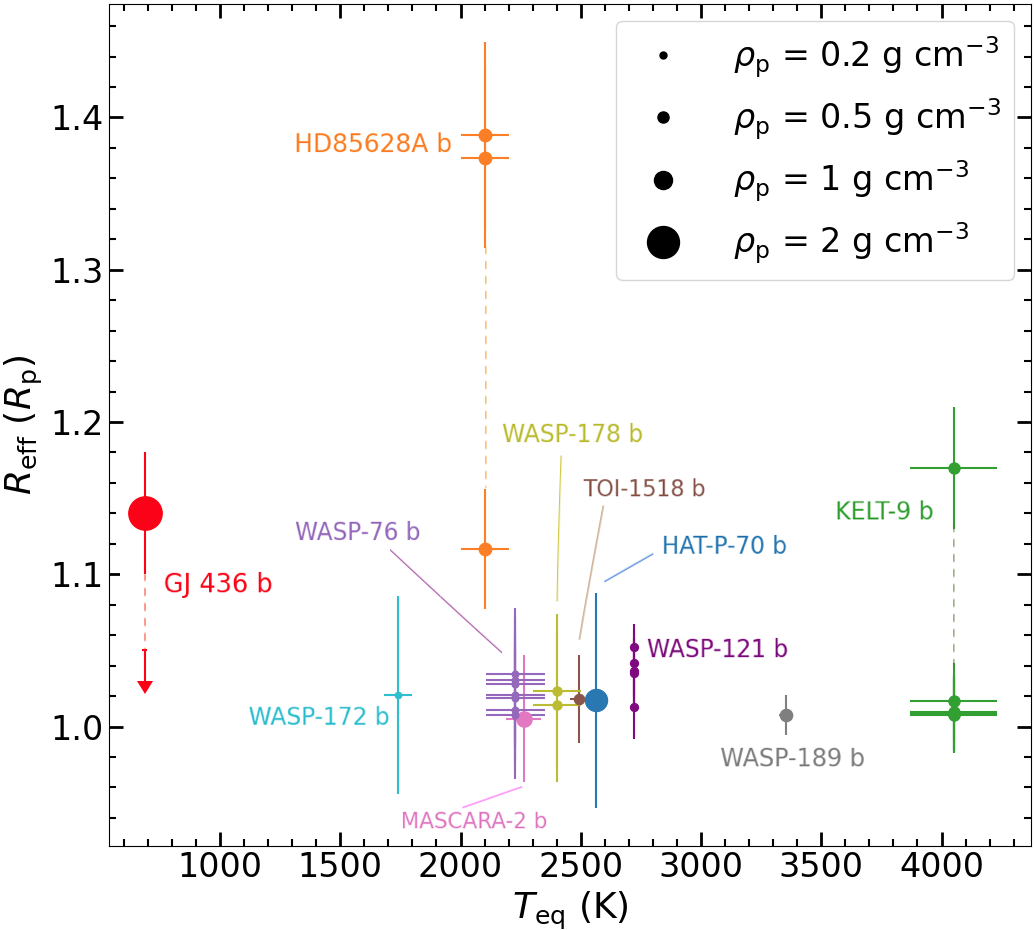}
    \caption{Effective radius versus equilibrium temperature for all eleven exoplanets with reported detections of neutral gaseous iron in their upper atmospheres via transmission spectroscopy, including our target GJ\,436\,b (red dot for T1; red arrow for T2). All planets are labelled, and symbol size scales with planetary bulk density, as indicated in the legend. Different measurements of the same planet are connected with dashed lines. References: GJ\,436\,b (this work), HAT-P-70\,b \citep{Bello-Arufe2022}, HD\,85628A\,b \citep{Zhang2022, Jiang2023}, KELT-9\,b \citep{Hoeijmakers2018, Hoeijmakers2019, Borsato2024, DArpa2024}, MASCARA-2\,b \citep{Nugroho2020}, TOI-1518\,b \citep{Cabot2021}, WASP-76\,b \citep{Tabernero2021, Kesseli2021, Kesseli2022, AzevedoSilva2022}, WASP-121\,b \citep{Hoeijmakers2020, Ben-Yami2020, Borsa2021, AzevedoSilva2022}, WASP-172\,b \citep{Seidel2023}, WASP-178\,b \citep{Damasceno2024}, and WASP-189\,b \citep{Prinoth2022}.}
    \label{fig:Fe_detections}
    \end{figure}

    The S/N of the CCF signal potentially attributed to planetary Fe\,{\sc i} is maximised at $\sim$1300 K, roughly twice the planetary equilibrium temperature. If this reflects the conditions in the layers probed by ESPRESSO, it would imply a temperature inversion in GJ\,436\,b's atmosphere. Recent observations by \citet{Finnerty2026} in the 2.91--3.85~$\mu\mathrm{m}$ range suggest a $\sim$2000 K thermal inversion above a haze layer, with an inferred equilibrium temperature of $1010^{+540}_{-410}$ K for GJ\,436\,b. The derived value of 1300 K is consistent within $1\sigma$ and may indicate moderately higher temperatures than those obtained for the planet's dayside (660 K) from mid-infrared \textit{Spitzer} and JWST observations \citep[e.g.][]{Madhusudhan2011, Mukherjee2025}. The sub-Neptune GJ\,1214\,b orbits a host star similar to that of GJ\,436 and receives comparable irradiation, but has a lower mass \citep{Charbonneau2009}. \citet{Malsky2025} showed that, in the presence of soot hazes, thermal inversions of several hundred kelvin are required to reproduce the near- and mid-infrared observations of GJ\,1214\,b.

    Although thermal inversions in exoplanets have been typically reported in hot and ultra-hot Jupiters \citep[e.g.][]{Evans2017, Nugroho2017}, they could also occur in warm exoplanetary atmospheres if they contain chemical species that absorb visible or near-infrared stellar radiation. In hot and ultra-hot Jupiters, common inversion-causing molecules include AlO, VO, and TiO \citep{vonEssen2019, Piette2020}, but the species responsible for thermal inversions at cooler temperature might differ. For instance, thermal inversions are known to exist in the atmospheres of solar system planets with thick atmospheres, such as Jupiter and Earth, where they are induced by molecules like ozone and hydrocarbons \citep{Moses2005, Robinson&Catling2014}. In warm planetary atmospheres, the absorbing species can potentially include H\textsubscript{2}O, CO, CO\textsubscript{2}, CH\textsubscript{4} or SO\textsubscript{2} \citep[e.g.][]{Beatty2024, Welbanks2024}. Given that GJ\,436\,b might have an atmosphere rich in CO and CO\textsubscript{2} (see Sect.~\ref{section:Introduction}), such absorbers could, in principle, modify the vertical temperature structure.

    From Table~\ref{table:results_Fe_detection}, one striking result is the high blue-shifted velocity of the potential planet signature  ($-$18.6 km\,s$^{-1}$). This Doppler shift, commonly referred to as $v_{\mathrm{wind}}$, is typically interpreted as the signature of atmospheric dynamics such as day-to-night winds or equatorial jets across the terminator region \citep[e.g.][]{Snellen2010, Brogi2016, Hoeijmakers2018, Casasayas-Barris2019, Ehrenreich2020, Seidel2021, Wardenier2024}. Assuming a hydrogen-dominated atmosphere and using Eq.~1 of \citet{PaiAsnodkar2022}, we estimated sound speeds of $\sim$2.4 km s$^{-1}$ and $\sim$3.3 km s$^{-1}$ for atmospheric temperatures of 700 K and 1300 K, respectively. Supersonic winds or jets have been detected in upper atmospheres of highly irradiated exoplanets \citep[e.g.][]{PaiAsnodkar2022, Nortmann2025}. However, the observed velocity of the tentative signal in GJ\,436\,b is highly supersonic and greater than the typical values reported for hot and warm gas giants, usually between $-$5 and $-$10 km s$^{-1}$ \citep[e.g.][]{Casasayas-Barris2019, Allart2020, Tabernero2021, Cabot2021, AzevedoSilva2022}. Global circulation models predict that cooler atmospheres show lower wind speeds than hot Jupiters. In this context, the high wind velocity we measured is unexpected and likely unphysical, pointing to a non-planetary origin of the signal. However, the tentative water vapor and methane signals reported for GJ\,436\,b by \citet{Finnerty2026} imply a wind velocity of $-32.5^{+8.6}_{-4.4}$ km\,s$^{-1}$, although the authors caution that their result is affected by a $K_{\rm p}$--$v_{\rm sys}$ degeneracy. Such high velocities could be associated with strong atmospheric escape \citep{Ehrenreich2015}, but these processes typically occur in the extended exosphere at much lower densities than the pressure levels probed by optical transmission spectroscopy.

    We investigated whether planetary rotation could contribute to the observed blueshift. While planetary rotation would induce a symmetric effect if absorption from the evening and morning limbs were comparable, a blue-shifted signature could arise if the signal is dominated by absorption from the evening limb. Assuming a spin-orbit resonance close to synchronous rotation, due to the short orbital period and modest eccentricity of the planet \citep{Barnes2017}, we calculated an estimated rotational velocity of $\sim$0.7 km\,s$^{-1}$. \citet{Finnerty2026} derived a higher rotational velocity of $3.0^{+2.7}_{-1.9}$ km\,s$^{-1}$, consistent with our estimate within uncertainties. Orbital smearing during the 300 s integrations contributes an additional $\sim$0.7 km\,s$^{-1}$. Combined, these effects remain insufficient to account for the observed large blueshift.

\subsection{Night to night variations}

    The tentative planetary signal identified in T1 is not seen during the second transit observations. Table~\ref{table:results_Fe_detection} provides the Fe\,{\sc i} 1$\sigma$ upper limit for T2 computed as described in Sect.~\ref{section:Upper limits of the non detections}. Figure~\ref{fig:Fe_1300_T1_T2_combined} provides a visual comparison of the CCFs for T1, T2, and for the combined transmission spectrum of both nights. Assuming negligible planetary or stellar variability between the two epochs, a similar signal would be expected on both nights, particularly given the slightly higher S/N of the ESPRESSO data for T2 (Fig.~\ref{fig:airmass_and_SN}). The non-detection during the second transit therefore suggests that the tentative Fe\,{\sc i} signal in T1 is likely spurious. Nevertheless, GJ\,436 is active and shows frequent flaring. Such stellar activity can enhance photochemical reactions, contributing to haze formation \citep{Morley2015}, and modify the atmospheric composition of close-in planets, inducing time-dependent variations in transmission spectra \citep{Konings2022}, and directly affecting the depth of spectral features. Although the detected signal is only tentative, it is still instructive to explore physically plausible scenarios that could account for the observed T1--T2 differences, with potential applicability to other planetary systems.

\subsubsection{Planetary variability, clouds, and hazes}

    Intrinsic variability could provide crucial information about planetary atmospheres and their evolution. Contrasting results have been reported in various studies examining the atmosphere of GJ\,436\,b (Sect.~\ref{section:Introduction}). While \citet{Beaulieu2011} claimed the possible detection of methane in GJ\,436\,b using mid-infrared \textit{Spitzer} data, the recent work by \citet{Grasser2024} informed that no transmission signals of water, carbon monoxide and methane were detected at near-infrared wavelengths using CRIRES+. The latter authors argued that the presence of a high-altitude cloud deck or a high atmospheric metallicity could explain the absence of these molecular species. This interpretation is consistent with the recent study of \citet{Pelaez-Torres2026}, who combined multiple CARMENES and CRIRES+ transits of GJ\,436\,b, reporting no molecular detections. Their results indicate that the atmosphere is either cloudy at high altitudes or highly metal enriched, both scenarios leading to muted spectral features. In contrast, \citet{Finnerty2026} reported tentative detection of H$_2$O and CH$_4$ emission using Keck data and the CCF analysis technique. These contrasting results illustrate the intrinsic difficulty of probing cool exoplanetary atmospheres and may reflect a combination of observational limitations, methodological differences, and potential atmospheric variability.

    \citet{Morley2015} modelled atmospheres of super-Earth and sub-Neptune planets incorporating thick clouds and hazes, showing that photochemical hazes with varying particle sizes can yield featureless transmission spectra in high metallicity environments. Specifically, they indicated that temperate planets with temperatures between 600–900 K could exhibit significant methane-derived photochemical hazes. The location and properties of the clouds of condensates and hazes may change significantly at different time scales depending on various intrinsic and external factors. Brown dwarfs are generally considered a proxy of temperate and hot exoplanets. \citet{Metchev15} discussed that photospheric heterogeneities are present on virtually all brown dwarfs with temperatures typical of ultra-hot, hot, and warm exoplanets. A third of brown dwarfs show irregular light curves, indicating that they have multiple spots that evolve on short time scales. The binary brown dwarf WISE J104915.57$-$531906.1\,AB, with temperatures close to that of GJ\,436\,b, are known to vary tremendously (1--11\,\%) at all observed optical, near- and mid-infrared wavelengths up to 10 $\mu$m as observed by JWST and ground-based facilities \citep{Buenzli15, Biller13, Biller24}. Variations occur at time scales of hours. They are complex, wavelength-dependent and cannot easily be described by any of the current theories: variability driven by cloud patchiness associated with deep silicate and iron clouds \citep{Luna2021}, atmospheric general circulation driven by cloud radiative feedback \citep{Tan2021}, and changes driven by hotspots and chemical differences produced by non-equilibrium chemistry \citep{Tremblin2020}. Furthermore, variability amplitudes can change significantly from one rotation period to another \citep[e.g.][]{apai21}, indicating that these cool atmospheres can have dramatic evolution over any time scale. Variability changes in the atmospheric properties could have an impact on the detectability of species, either enhancing or completely hiding the signals. However, a 10\,\%~change in the planetary flux is not sufficient to explain the non-detection of neutral iron in T2. Based on the noise measured in the planetary spectra of both observing nights, a much larger variability ($>$50\,\%) would be needed to significantly mute the tentative Fe\,{\sc i} signal of T1.

\subsubsection{Ionisation of iron}

    The levels of chromospheric emission of the host star were higher during the second planetary transit, suggesting the presence of more high-energy photons at short wavelengths. High-energy irradiation drives atomic ionization, leading to depletion of neutral species in the planetary upper atmosphere and reducing the detectability of absorption signals during T2. Ionised atoms, including Fe\,{\sc ii}, have been detected in a few ultra-hot and hot Jupiters \citep[e.g.][]{Casasayas-Barris2019, Cubillos2020, Bello-Arufe2022}. To address this scenario, and driven by the high temperatures ($>$2000 K) that the planetary upper atmosphere may reach \citep{Finnerty2026}, we searched for Fe\,{\sc ii} in the planetary spectra of GJ\,436\,b from both observing nights using the CCF technique and synthetic models with a temperature of 3000 K (Sect.~\ref{section:Cross-correlation analysis}). The 1$\sigma$ upper limit on the detection is reported in Table~\ref{table:CCF_non_detections}. 

    \begin{figure}
    \centering
    \includegraphics[width=0.9\hsize]{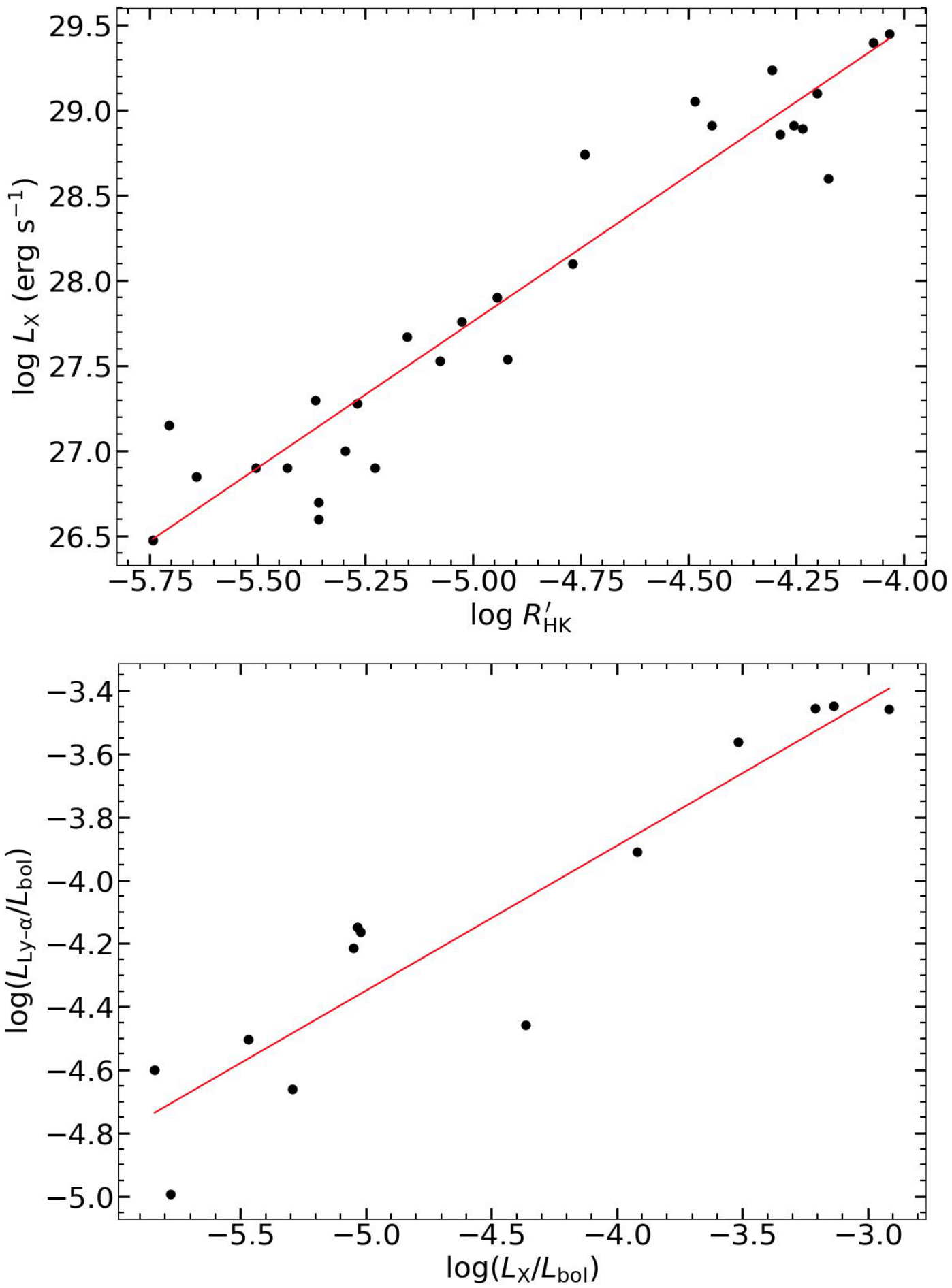}
    \caption{\textit{Top}: $\log L_{\text{X}}$ versus $ \log R'_{\text{HK}}$ activity index for all M3 stars listed in \citet{Houdebine2017}. A fitted linear function is shown as a red line. \textit{Bottom}: $\log(L_{\text{Ly-$\alpha$}}/L_{\text{bol}})$ versus $\log(L_{\text{X}}/L_{\text{bol}})$ for M2.5--M3.5 stars compiled in \citet{Linsky2020}. The linear function fitted to the data is depicted as a red line.}
    \label{fig:Linsky_Houdebine}
    \end{figure}

    We estimated the variation in the iron ionisation fraction between both nights. For this, we considered photodissociation processes, assuming them to be primarily driven by Ly-$\alpha$ photons (1216 \r{A}), as the Fe\,{\sc i} ionisation energy corresponds to 1569 \r{A}. To estimate the stellar Ly-$\alpha$ radiation received by the planet, we first established a relationship between the $\log R'_{\text{HK}}$ index and the X-ray luminosity for M3 stars, based on 29 M3 stars from \citet{Houdebine2017}. We applied a linear fit to the data (top panel of Fig.~\ref{fig:Linsky_Houdebine}) described by the following equation:
   
    \begin{equation}    
        \log L_{\text{X}} = (1.73 \pm 0.11) \cdot \log R'_{\text{HK}} + (36.38 \pm 0.53) ,
        \label{eq:Houdebine2017_M3}
    \end{equation}

    \noindent where the stellar X-ray luminosity, $L_{\text{X}}$, is expressed in $\mathrm{erg}\, \mathrm{s}^{-1}$.

    We then correlated the X-ray luminosity with the Ly-$\alpha$ luminosity for M3 stars using 13 stars between M2.5 and M3.5 listed in \citet{Linsky2020}, and comparing their Ly-$\alpha$ and X-ray to bolometric luminosity ratios. Applying a linear fit to the data (bottom panel of Fig.~\ref{fig:Linsky_Houdebine}) we derived the following equation:

    \begin{equation}    
        \log(L_{\text{Ly-$\alpha$}}/L_*) = (0.480 \pm 0.066) \cdot \log(L_{\text{X}}/L_{\text{bol}}) + (0.305 \pm 0.071) 
        \label{eq:Linsky2020_M3}
    .\end{equation}

    We measured the Ly-$\alpha$ flux received by the planet for each observing night using Eqs. \eqref{eq:Houdebine2017_M3} and \eqref{eq:Linsky2020_M3}. The stellar bolometric luminosity and the orbital semi-major axis are given in Table~\ref{table:system}; the $\log  R'_{\text{HK}}$ values for each night were calculated by averaging the indices outside of flare periods, as provided by the ESPRESSO DRS \citep[see our Fig.~\ref{fig:indices};][]{Pepe2021}. The resulting Ly-$\alpha$ flux incident on the planet is $(1.91 \pm 1.25) \times 10^{3} \, \text{erg}\, \text{s}^{-1} \text{cm}^{-2}$ for T1 and $(2.40 \pm 1.53) \times 10^{3} \, \text{erg}\, \text{s}^{-1} \text{cm}^{-2}$ for T2. The errors were determined via error propagation. These values were then used to estimate the ionisation fraction due to photoionisation driven by Ly-$\alpha$ photons, employing the following relation:

    \begin{equation}    
        \frac{n_{\text{Fe\,{\sc ii}}}}{n_{\text{Fe\,{\sc i}}}} = \frac{k_{\text{ion}}}{n_e \alpha_{\text{r}}} \;\; ,
        \label{eq:ion_fraction}
    \end{equation}

    \noindent where $n_e$ is the electron density, $k_{\text{ion}}$ is the ionisation rate, $\alpha_{\text{r}}$ is the radiative recombination coefficient. We calculated the former as

    \begin{equation}    
        k_{\text{ion}} = \sigma_{\text{Fe\,{\sc i}}} \cdot \phi \;\; ,
        \label{eq:ion_rate}
    \end{equation}

    \noindent where the Ly-$\alpha$ ionisation cross-section for Fe\,{\sc i}, $\sigma_{\text{Fe\,{\sc i}}}$, is $4.86 \times 10^{-18} \, \text{cm}^2$ \citep{Heays2017}. The Ly-$\alpha$ photon flux, $\phi$, is calculated as $F_{\text{Ly-$\alpha$}} / E_{\text{Ly-$\alpha$}}$, where the energy of the Ly-$\alpha$ photons, $E_{\text{Ly-$\alpha$}}$, is 1216 \r{A} (10.2 eV), and $F_{\text{Ly-$\alpha$}}$ is the incident Ly-$\alpha$ flux for each night. The radiative recombination coefficient is calculated using the expression derived in \citet{Woods&Shull1981}:

    \begin{equation}    
        \alpha_{\text{r}} = A_{\text{r}} \, (T / 10^{4} \, \text{K})^{-\eta} \;\; ,
        \label{eq:recombination_coefficient}
    \end{equation}

    \noindent where $T$ is the kinetic temperature of the gas and, for the recombination from Fe\,{\sc ii} to Fe\,{\sc i}, the corresponding coefficients are $A_{\text{r}} = 1.42 \times 10^{3} \text{cm}^3\text{s}^{-1}$ and $\eta = 0.891$ \citep{Woods&Shull1981}.
    
    Using Eqs. \eqref{eq:ion_fraction}, \eqref{eq:ion_rate}, and \eqref{eq:recombination_coefficient}, the parameters that could have varied between the two nights observed with ESPRESSO are $F_{\text{Ly-$\alpha$}}$, $n_e$, and $T$. Assuming that $n_e$ and $T$ remained unchanged on both nights, then $\frac{n_{\text{Fe\,{\sc ii}}}}{n_{\text{Fe\,{\sc i}}}}$ would be directly proportional to the Ly-$\alpha$ flux ratio between the two epochs. This suggests that Fe\,{\sc i} abundance was $\sim$25\,\%~lower during T2 that during T1. 
    
    Using the synthetic Fe\,{\sc i} spectra generated in Sect.~\ref{section:Fe_Na_detectability}, we found that a 25\,\%~decrease in neutral iron abundance produces only minor variations ($\le$10\,\%) in the S/N value of the potential Fe\,{\sc i} signal for bulk metallicities [Fe/H] $\le$ 0.5 dex, with a larger impact at higher abundances. These simulations indicate that a 25\,\%~depletion in Fe\,{\sc i} abundance has little impact on the detectability of Fe\,{\sc i} at moderate iron abundances.

\subsubsection{Spectral veiling}

    Another scenario we explored is veiling of the upper planetary atmosphere. By receiving more energetic photons from the star during T2, the tenuous layers of the escaping planetary atmosphere might have become hotter and more luminous, acting as a source of continuum flux and causing the spectral lines to appear shallower and thus undetectable. To investigate how much veiling would be needed to obscure the tentative Fe\,{\sc i} absorption signal observed in T1, we simulated spectra with various veiling factors, $r$, using the conversion $F_{\mathrm{v}} = \dfrac{F_{\mathrm{T1}} + r}{1 + r}$, where $F_{\mathrm{T1}}$ is the planetary spectrum from T1, and $F_{\mathrm{v}}$ represents the simulated veiled planetary spectrum. We assumed constant veiling factors across all ESPRESSO wavelengths, exploring $r$ values from 0 to 1, in steps of 0.1. To align the noise levels of the veiled spectra with those from the T1 planetary spectrum, we introduced additional Gaussian noise with an amplitude of $\sqrt{\sigma_{\text{orig}}^2 - \sigma_{\text{mod}}^2}$, where $\sigma_{\text{orig}}$ and $\sigma_{\text{mod}}$ are the noise levels of the original and each modified spectrum, respectively. This adjustment was applied separately for different wavelength regions, in sections of 200 \r{A}. Subsequently, we computed Fe\,{\sc i} CCFs for the veiled spectra using the same methods outlined in Sect.~\ref{section:Cross-correlation analysis}. The S/N of the features at the same location as the potential planet signal was then measured, as previously described in Sect.~\ref{Marginal evidence for Fe?}. The results, displayed in Fig.~\ref{fig:SN_veiling}, indicate that the S/N decreases as $r$ increases. When setting the detectability threshold at S/N = 2, no signal was recovered for $r > 0.3$.

    \begin{figure}
    \centering
    \includegraphics[width=0.85\hsize]{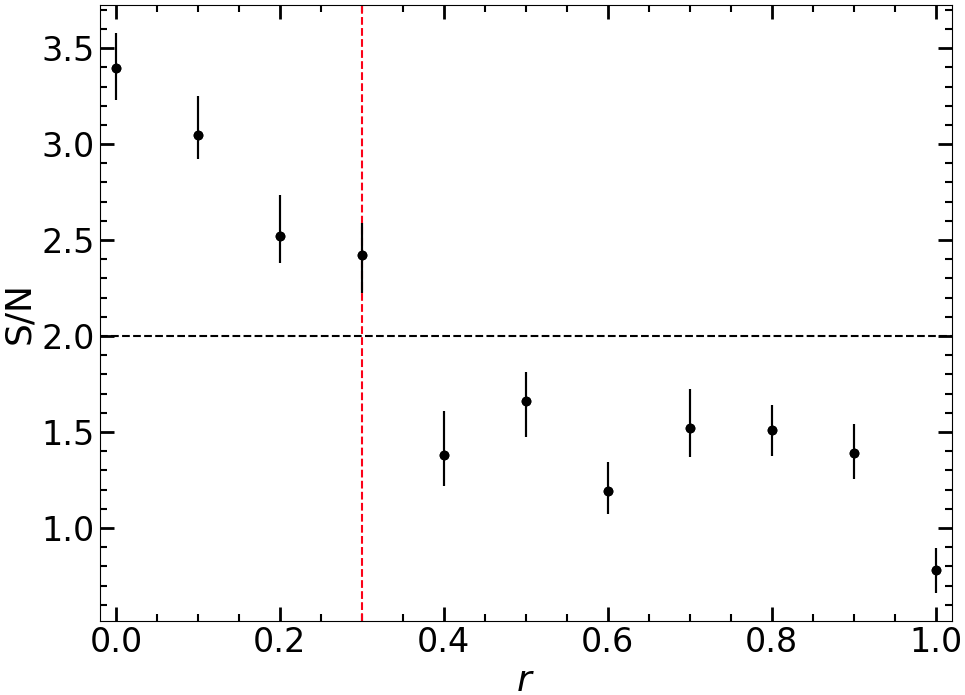}
    \caption{S/N of the Fe\,{\sc i} absorption signal in the CCFs computed for the T1 planetary spectrum with varying veiling levels, ranging from $r$ = 0 (no veiling) to $r$ = 1. The dashed black line indicates the reference 2$\sigma$, set as the detectability threshold for the purpose of this diagram, while the dashed red line marks a veiling factor of $r > 0.3$, beyond which the planetary signal is not recovered.}
    \label{fig:SN_veiling}
    \end{figure}

    These results suggest that a moderate spectral veiling of $\sim 30\%$, that is, an addition of $\sim 30\%$ continuum flux to the planetary spectrum, can significantly reduce the sensitivity of our method to detect atmospheric species. Coincidentally, we calculated an increase in the mean stellar fluxes, excluding flares, of 14\,\%~and 34\,\%~for Na\,{\sc i} and Ca\,{\sc ii} H \& K, respectively, from T1 to T2 (see Sect.~\ref{section:Stellar activity}). This increase in the stellar emission during T2, especially in the blue range, where many Fe\,{\sc i} lines are located, may account for the suppression of any signal during the second transit observations.

\section{Conclusions}
\label{section:Conclusions}

    High spectral resolution ($R \approx 138,000$) optical observations of the warm Neptune GJ\,436\,b were conducted on two separate nights with primary planetary transits (two months apart) using the ESPRESSO spectrograph at the VLT. The data of the two observing nights were of a similar quality and were affected by low-energy stellar flares occurring immediately after the planetary transit events. From the transmission planetary spectrum, strong and broad stellar residuals around the H$\alpha$ line, correlated with the stellar flare, were observed in the second night, but not in the first night. We did not detect the presence of the following atomic and molecular species in GJ\,436\,b's optical spectrum: H\,{\sc i} (H$\alpha$), Na\,{\sc i}, Mg\,{\sc i}, Li\,{\sc i}, V\,{\sc i}, Cr\,{\sc i}, Fe\,{\sc ii}, TiO, and VO. Thus, we were able to place upper limits on their detection in the planetary atmosphere.

    The cross-correlation of the T1 planetary spectrum with synthetic templates resulted in a tentative Fe\,{\sc i} signal with a low statistical significance (S/N = 3.4 $\pm$ 0.2) and a wind velocity of $-18.6$ km\,s$^{-1}$, which deviates from the expected planetary rest frame and exceeds typical values reported for warmer, more massive planets. No Fe\,{\sc i} signal was detected in T2. The lack of reproducibility and low S/N suggest a spurious detection, consistent with a featureless optical transmission spectrum. Nevertheless, if confirmed by future observations, such a signal would imply atmospheric temperatures above the planetary equilibrium value, allowing iron to remain in gaseous form, along with an iron abundance comparable to or exceeding that of the host star. While statistical noise can account for the T1--T2 discrepancy, a scenario involving $\sim$30\,\%~spectral veiling driven by enhanced stellar activity could provide a plausible explanation for the suppression of planetary spectral features between transits. Multi-epoch observations across numerous transits would be needed to characterise the planetary atmosphere and assess the impact of stellar activity.

\begin{acknowledgements}
    E.H.-C. acknowledges support from grant PRE2020-094770 under project PID2019-109522GB-C51 funded by the Spanish Ministry of Science, Innovation and Universities / State Agency of Research, MCIU/AEI/10.13039/501100011033, and by ‘ERDF, A way of making Europe’. E.H.-C., M.R.Z.O. and J.S.-F. acknowledge project PID2022-137241NB-C42 funded by MCIU/AEI/10.13039/501100011033. J.I.G.H. and A.S.M. acknowledge financial support from the Spanish Ministry of Science, Innovation and Universities (MICIU) projects PID2020-117493GB-I00 and PID2023-149982NB-I00. The work of C.J.M. was financed by Portuguese funds through FCT (Funda\c{c}\~ao para a Ci\^encia e a Tecnologia) in the framework of the project 2022.04048.PTDC (Phi in the Sky, DOI 10.54499/2022.04048.PTDC). C.J.M. also acknowledges FCT and POCH/FSE (EC) support through Investigador FCT Contract 2021.01214.CEECIND/CP1658/CT0001 (DOI 10.54499/2021.01214.CEECIND/CP1658/CT0001). A. R. C. S. acknowledges the support from Funda\c{c}ao para a Ci\^encia e a Tecnologia (FCT) through the fellowship 2021.07856.BD. Y.C.D. acknowledges support from the Funda\c{c}\~ao para a Ci\^encia e a Tecnologia (FCT) and POCH/FSE through the PhD fellowship 2024.01038.BD. E.P. acknowledges financial support from the Agencia Estatal de Investigaci\'on of the Ministerio de Ciencia e Innovaci\'on MCIN/AEI/10.13039/501100011033 and the ERDF “A way of making Europe” through project PID2021-125627OB-C32, and from the Centre of Excellence “Severo Ochoa” award to the Instituto de Astrof\'isica de Canarias. M.L. acknowledges support of the Swiss National Science Foundation under grant number PCEFP2\_194576. The contributions of M.L. and A.P. have been carried out within the framework of the NCCR PlanetS supported by the Swiss National Science Foundation under grants 51NF40\_182901 and 51NF40\_205606. A.P. acknowledges support from the Unidad de Excelencia Mar\'a de Maeztu CEX2020-001058-M programme and from the Generalitat de Catalunya/CERCA. This work was co-funded by the European Union (ERC, FIERCE, 101052347). Views and opinions expressed are however those of the authors only and do not necessarily reflect those of the European Union or the European Research Council. Neither the European Union nor the granting authority can be held responsible for them. This work was supported by FCT - Funda\c{c}\~ao para a Ci\^encia e a Tecnologia through national funds and by FEDER through COMPETE2020 - Programa Operacional Competitividade e Internacionaliza\c{c}\~ao by these grants: UID/FIS/04434/2019, UIDB/04434/2020, UIDP/04434/2020. R.A. acknowledges the Swiss National Science Foundation (SNSF) support under the Post-Doc Mobility grant P500PT\_222212 and the support of the Institut Trottier de Recherche sur les Exoplan\`etes (iREx). The authors acknowledge financial contribution from the European Union - Next Generation EU RRF M4C2 1.1 PRIN MUR 2022 project 2022CERJ49 (ESPLORA). The INAF authors acknowledge financial support of the Italian Ministry of Education, University, and Research with PRIN 201278X4FL and the "Progetti Premiali" funding scheme. F.P.E. would like to acknowledge the Swiss National Science Foundation (SNSF) for supporting research with ESPRESSO through the SNSF grants nr. 140649, 152721, 166227, 184618 and 215190. The ESPRESSO Instrument Project was partially funded through SNSF’s FLARE Programme for large infrastructures. This research has made use of the SIMBAD database, operated at CDS, Strasbourg, France. This research has made use of the NASA Exoplanet Archive, which is operated by the California Institute of Technology, under contract with the National Aeronautics and Space Administration under the Exoplanet Exploration Program. This research has made use of NASA's Astrophysics Data System Bibliographic Services. We acknowledge the use of the ExoAtmospheres database during the preparation of this work.

\end{acknowledgements}

\bibliographystyle{aa}
\bibliography{references}

\onecolumn

\begin{appendix}

\section{Narrow-band results}
\label{section:Narrow band results}

    \begin{figure}[h!]
    \centering
    \includegraphics[width=0.85\hsize]{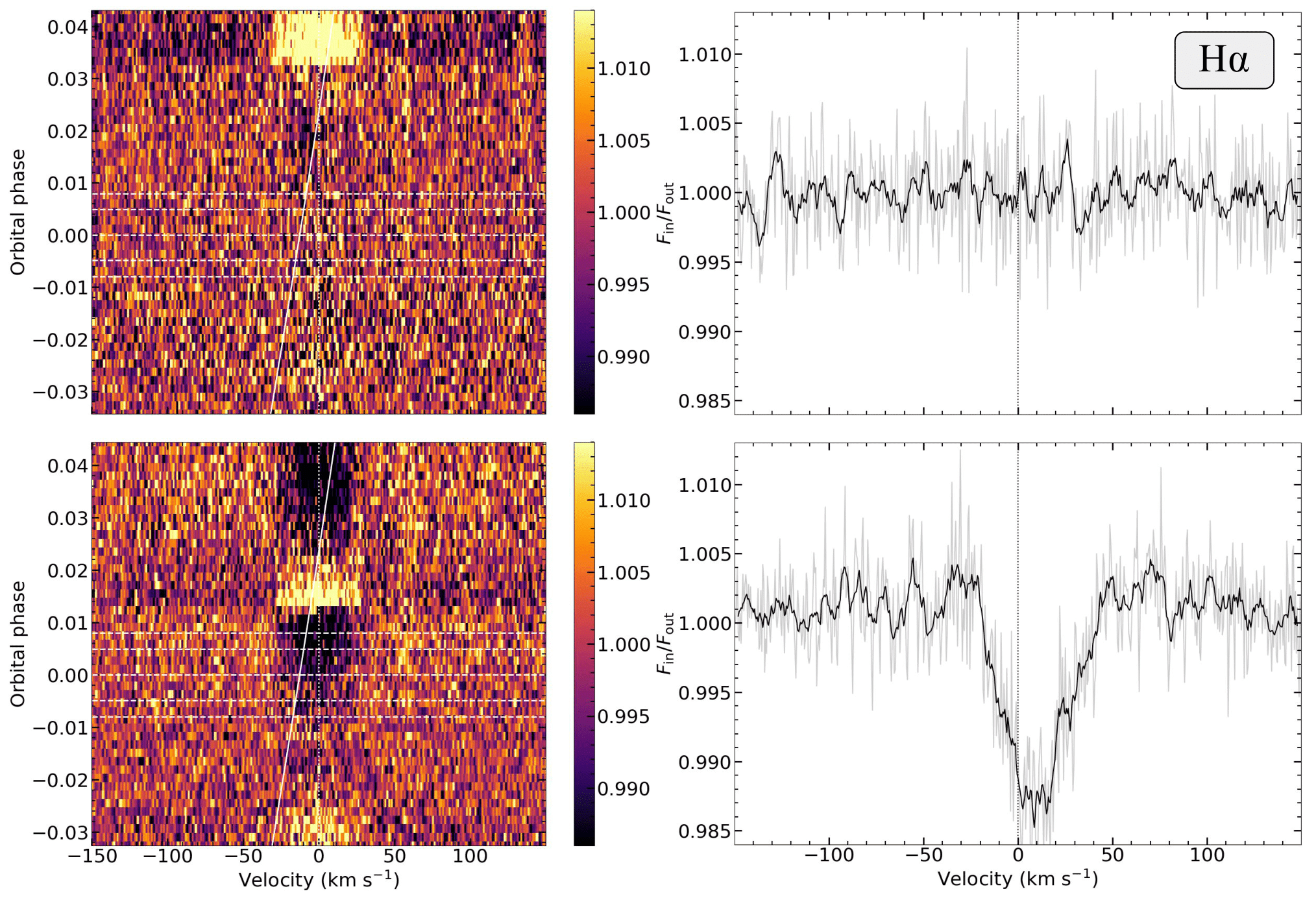}
    \caption{\textit{Left}: T1 (\textit{top}) and T2 (\textit{bottom}) tomography maps in the stellar rest frame around the H$\alpha$ line. A 3-pixel moving average was applied along the velocity axis to smooth the maps. The horizontal dashed lines indicate the orbital phases at the four contacts of the transit and at midtransit, and the slanted solid line marks the planetary velocities during the observation. The dotted vertical line indicates the stellar rest frame velocity. \textit{Right}: In-transit combined planetary spectrum around H$\alpha$ in the planetary rest frame. The grey line represents the original spectrum, while the black line represents the spectrum after applying a 9-point moving average for increasing the S/N. The central wavelength of H$\alpha$ is marked by the vertical line at 0 km s$^{-1}$. The observed H$\alpha$ absorption seen in the planetary spectrum of T2 is a product of stellar contamination (see Sect.~\ref{section:Transmission signal at H_alpha})}
    \label{fig:Halpha}
    \end{figure}

    \begin{figure}
    \centering
    \includegraphics[width=0.85\hsize]{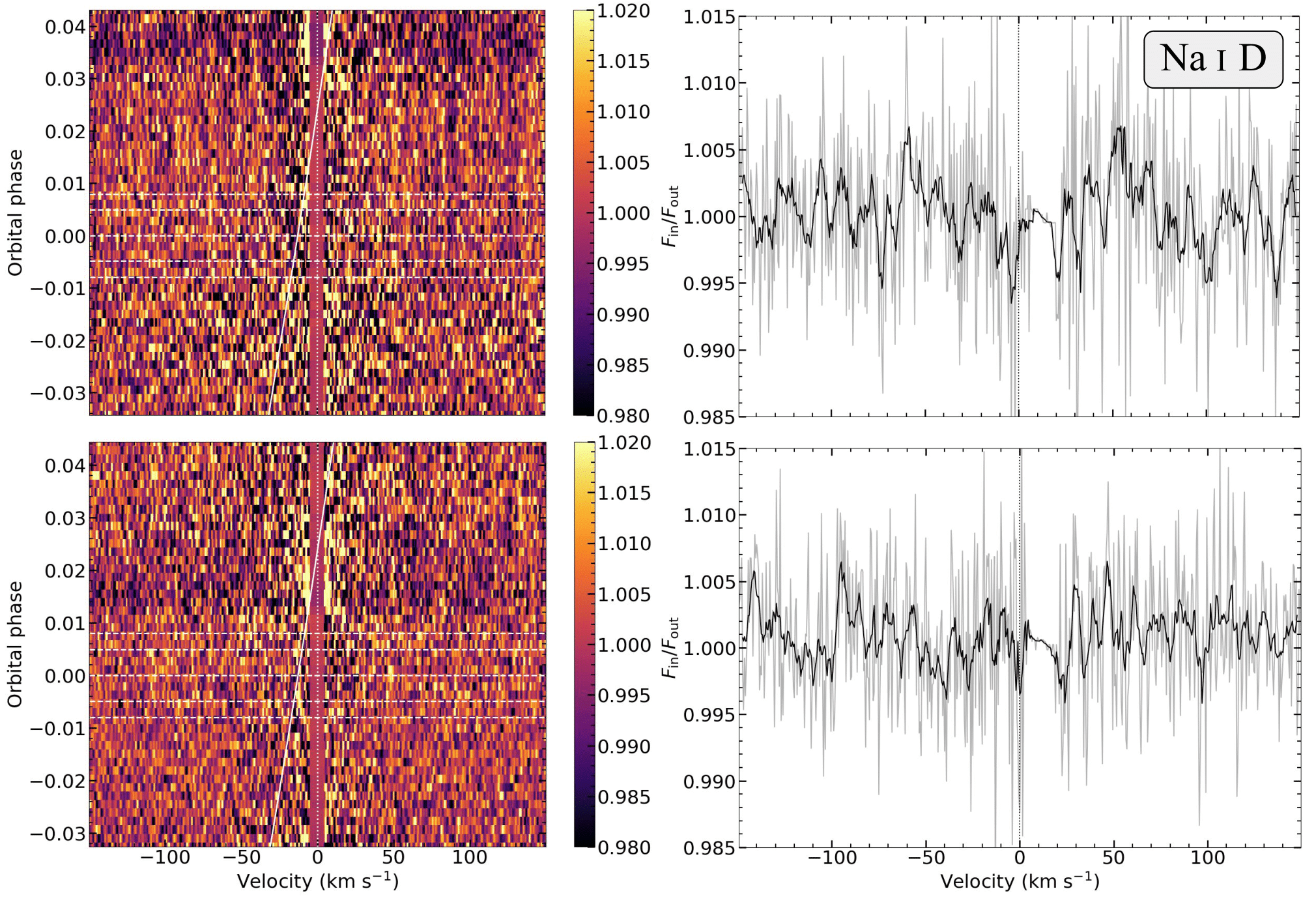}
    \caption{Same as Fig.~\ref{fig:Halpha} but for both lines of the Na\,{\sc i} D doublet combined. The region between $\pm$ 5 km s$^{-1}$ was masked out to remove stellar residuals.}
    \label{fig:Na}
    \end{figure}
    
    \begin{figure}
    \centering
    \includegraphics[width=0.85\hsize]{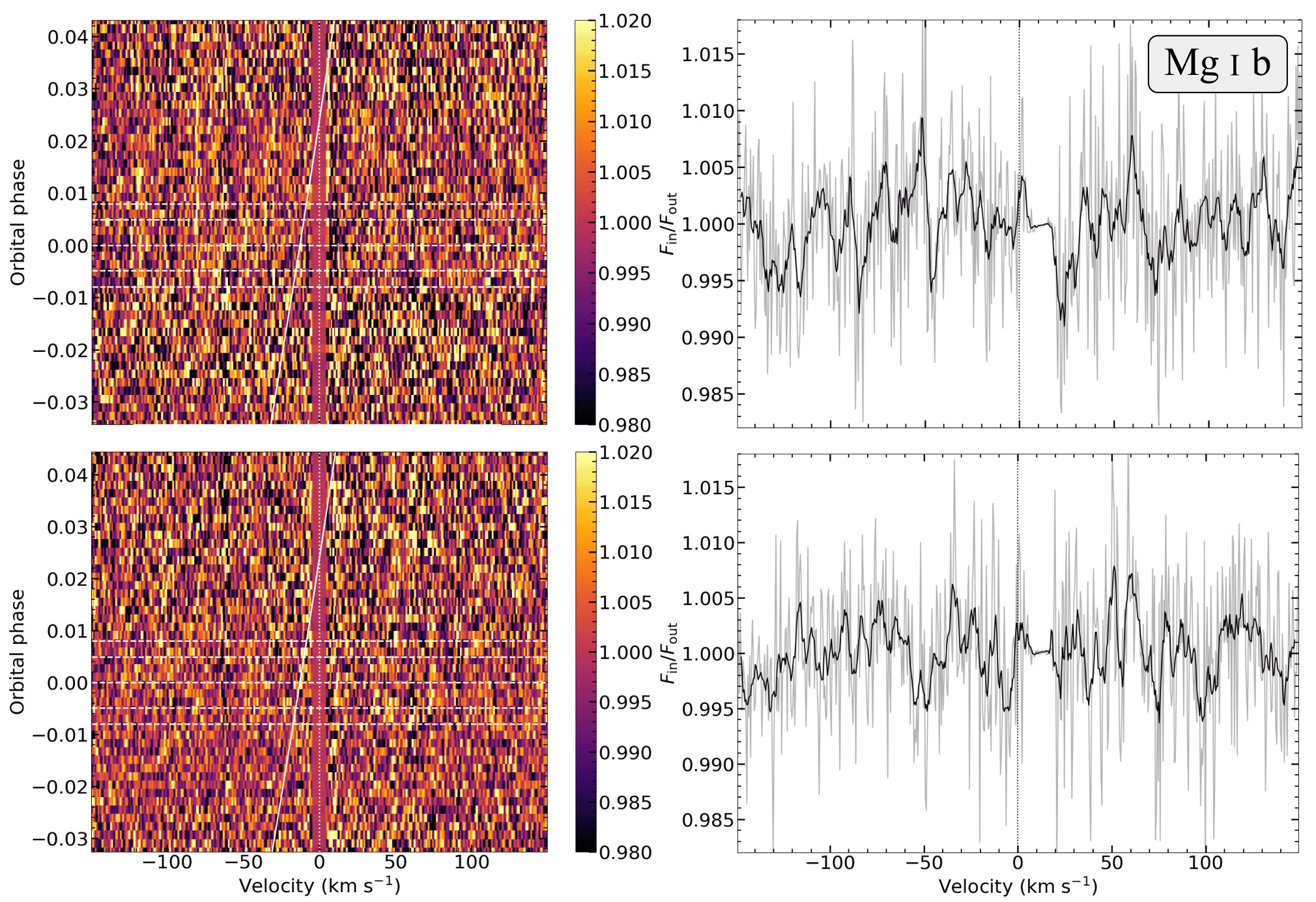}
    \caption{Same as Fig.~\ref{fig:Halpha} but for the three lines of the Mg\,{\sc i} b triplet combined. The region between $\pm$ 5 km s$^{-1}$ was masked out to remove stellar residuals.}
    \label{fig:Mg}
    \end{figure}

    \begin{figure}
    \centering
    \includegraphics[width=0.85\hsize]{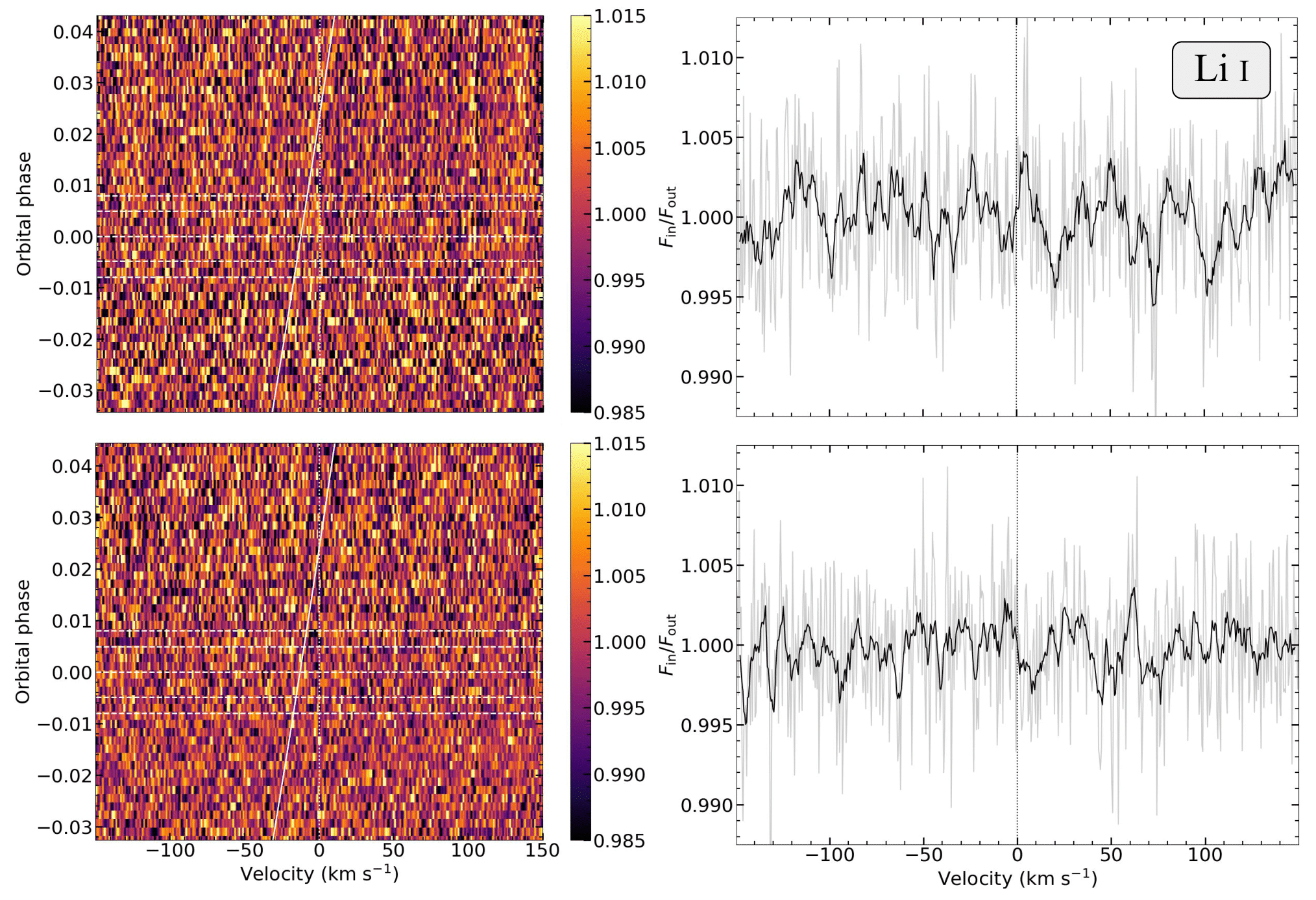}
    \caption{Same as Fig.~\ref{fig:Halpha} but for the Li\,{\sc i} line at 6707.761 \r{A}.}
    \label{fig:Li}
    \end{figure}

    \begin{figure}
    \centering
    \includegraphics[width=0.85\hsize]{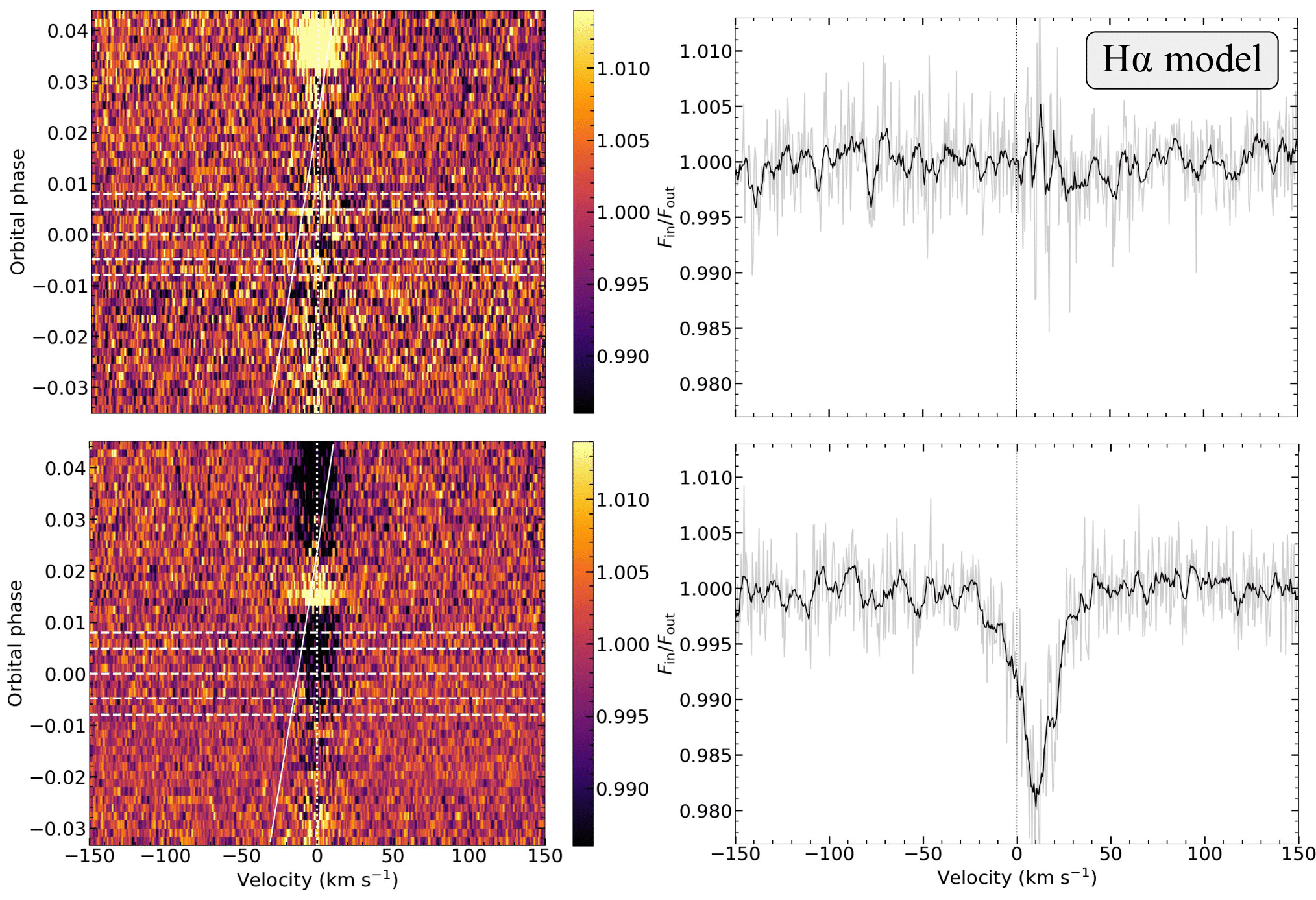}
    \caption{Same as Fig.~\ref{fig:Halpha} but for the modelled transmission spectra around the H$\alpha$ line.}
    \label{fig:Halpha_model}
    \end{figure}

    \begin{figure}
    \centering
    \includegraphics[width=\hsize]{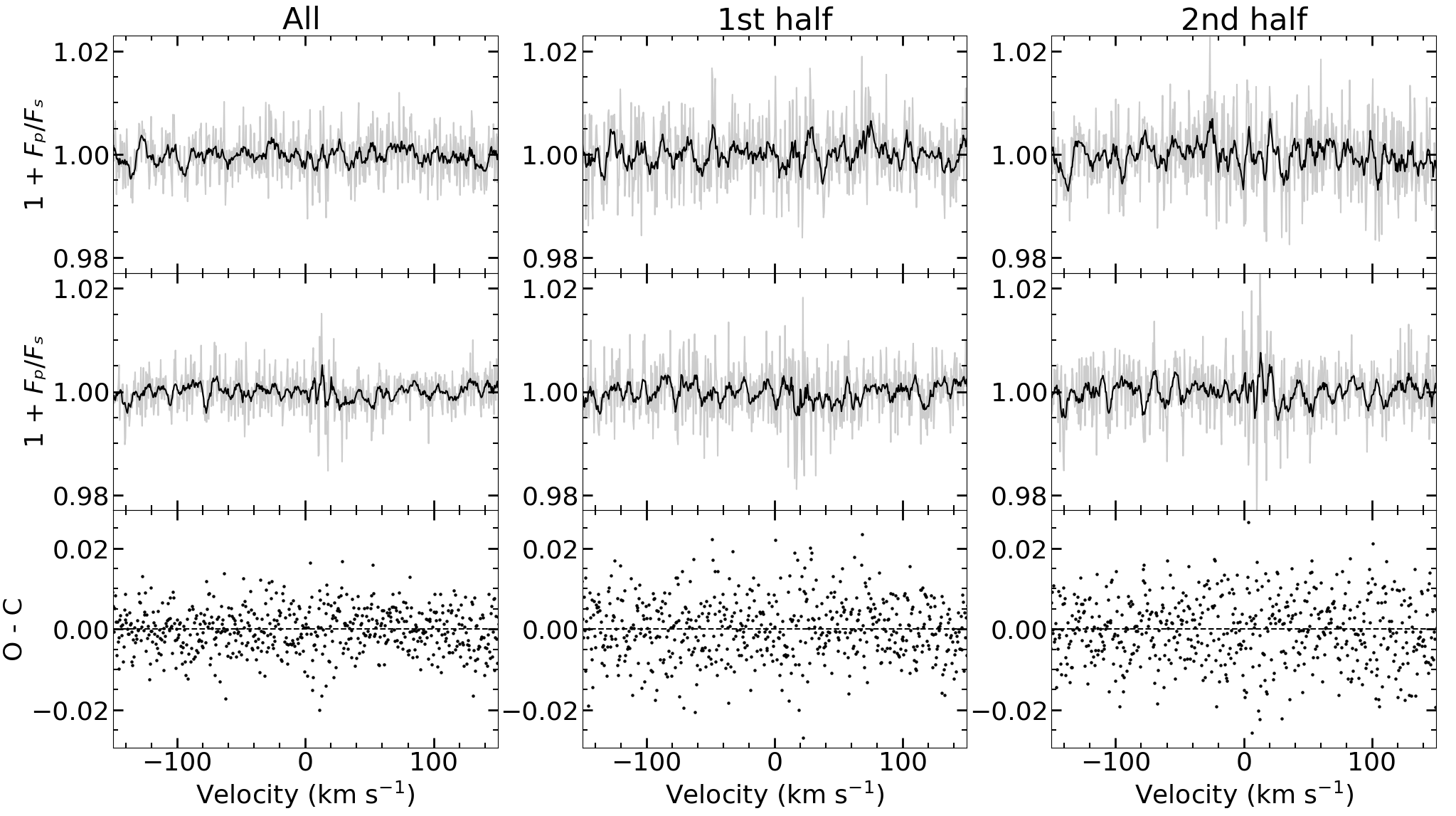}
    \caption{\textit{Top}: Observed in-transit transmission spectrum in the H$\alpha$ region for T1. The original data are shown in grey, while the smoothed version via a 9-point moving average is displayed in black. \textit{Middle}: Modelled in-transit transmission spectrum for T1. \textit{Bottom}: Residuals of the observed minus the model data. These results are displayed for all the in-transit data (\textit{left}), as well as separately for the first (\textit{centre}) and the second (\textit{right}) halves of the transit.}
    \label{fig:Halpha_obs_and_model_T1}
    \end{figure}

    \begin{figure}
    \centering
    \includegraphics[width=\hsize]{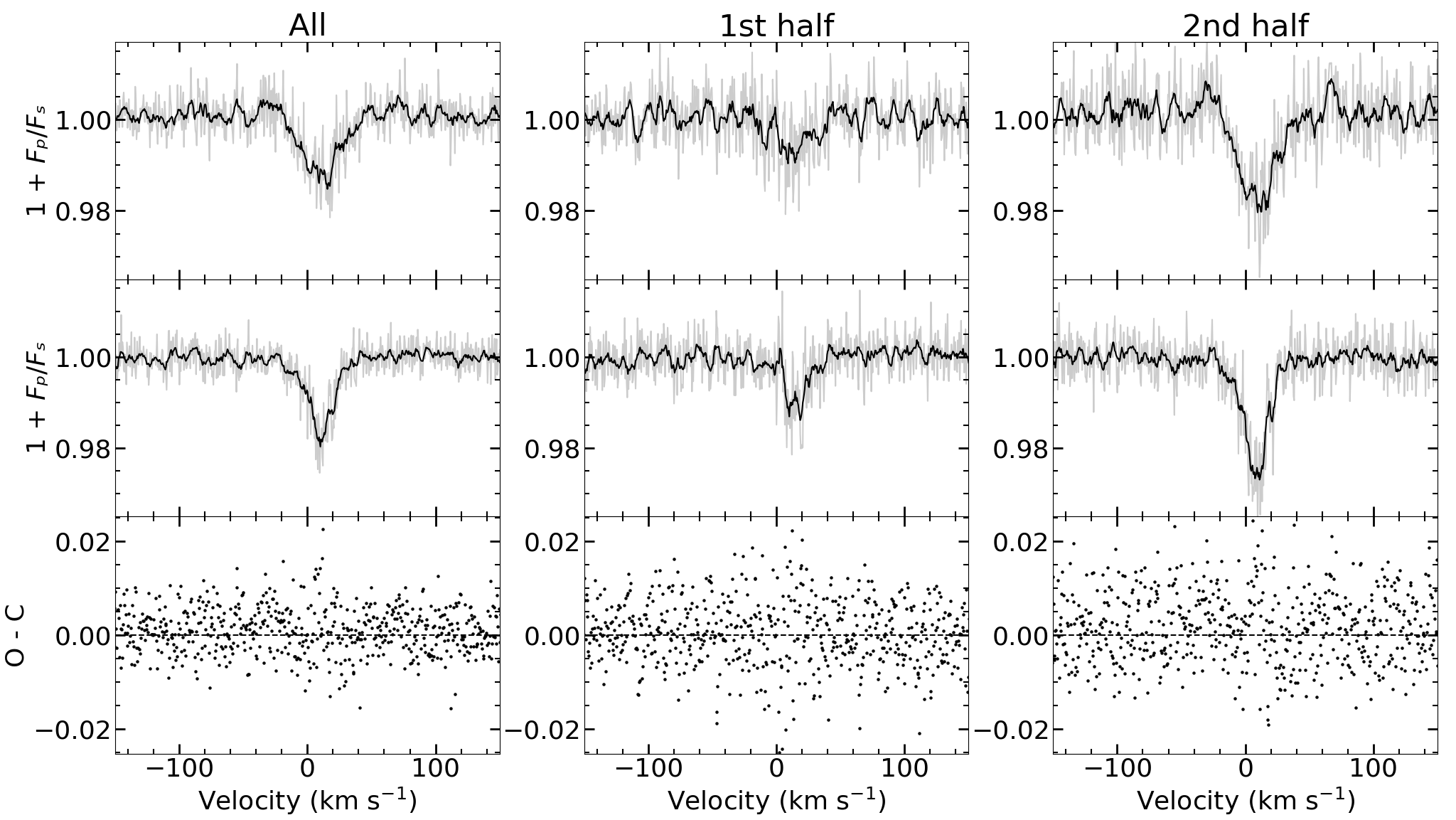}
    \caption{Same as Fig.~\ref{fig:Halpha_obs_and_model_T1} but for T2.}
    \label{fig:Halpha_obs_and_model_T2}
    \end{figure}

\onecolumn

\section{Synthetic spectra}
\label{section:Synthetic spectra}

    \begin{figure}[h!]
    \centering
    \includegraphics[width=0.61\hsize]{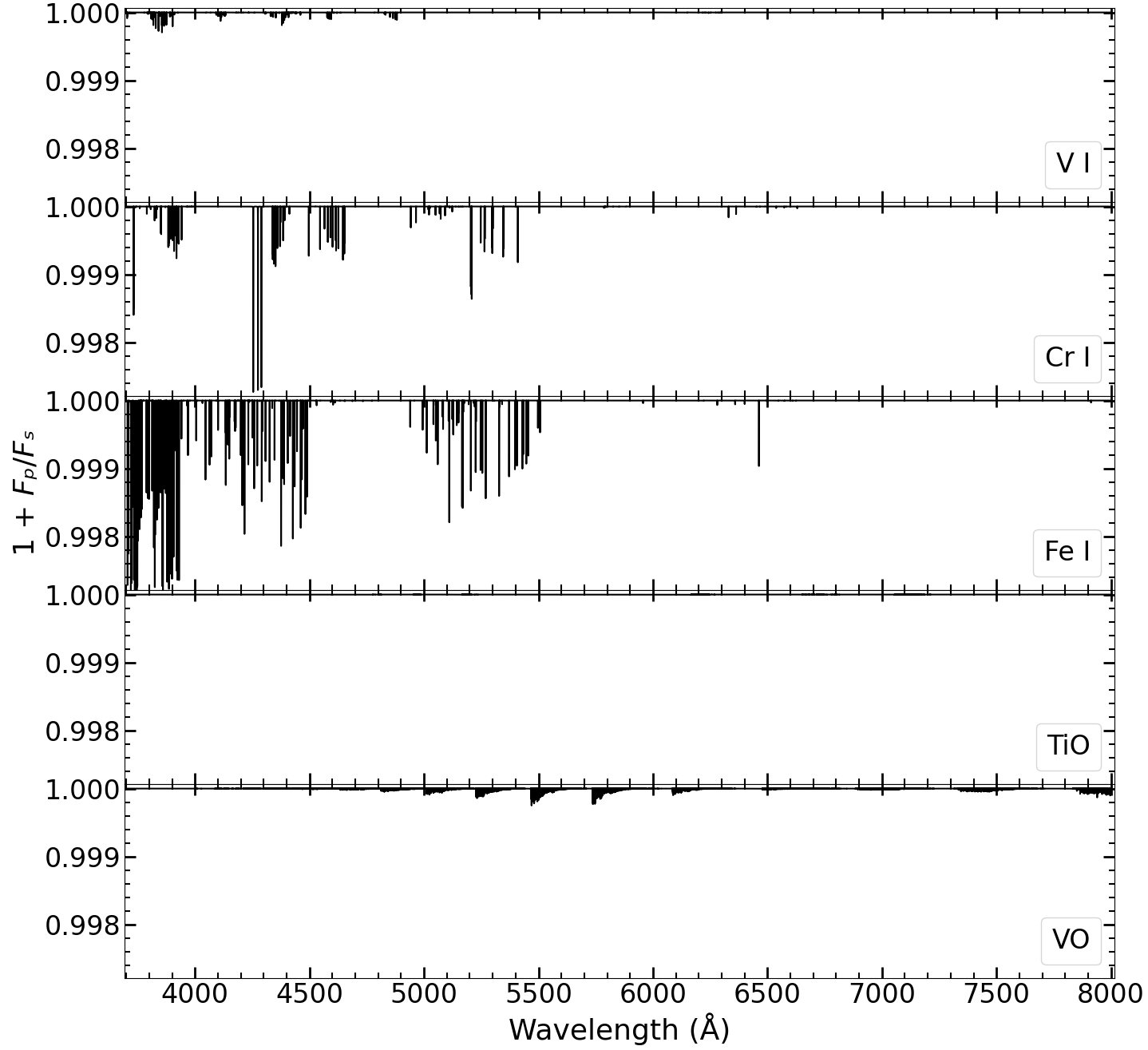}
    \caption{Synthetic \texttt{petitRADTRANS} transmission spectra computed for different species at 1300 K.}
    \label{fig:syn_spectra_1300K}
    \end{figure}

    \begin{figure}[h!]
    \centering
    \includegraphics[width=0.61\hsize]{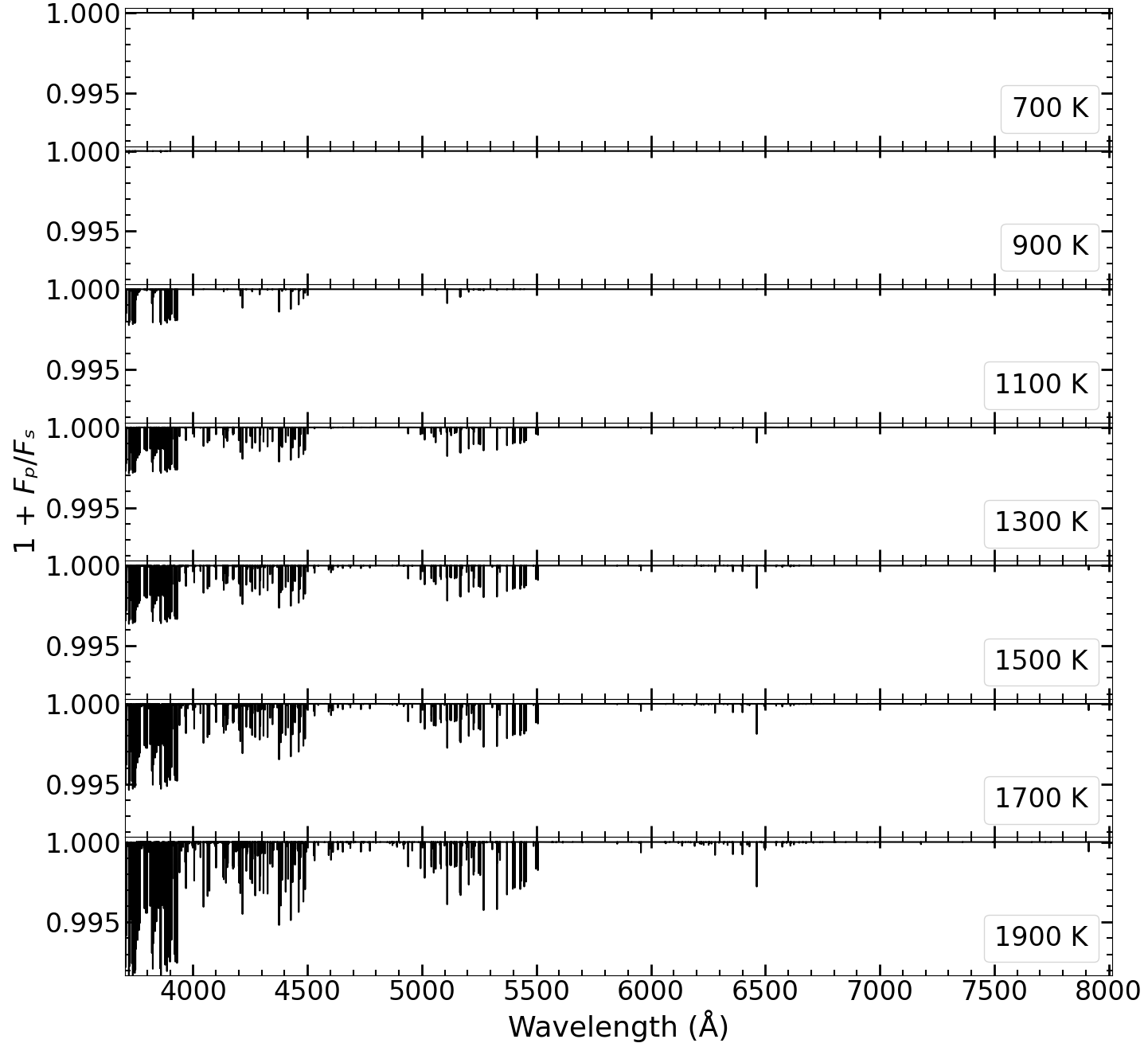}
    \caption{Synthetic \texttt{petitRADTRANS} transmission spectra computed for Fe\,{\sc i} at different temperatures. All spectra are plotted in the same y-axis scale. The lower-temperature models appear featureless due to their significantly weaker Fe\,{\sc i} absorption lines, with line depths below $10^{-4}$ at temperatures of 700--900 K.}
    \label{fig:syn_spectra_Fe_temperatures}
    \end{figure}

\onecolumn

\section{CCFs}
\label{section:CCFs}

    \begin{figure}[h!]
    \centering
    \includegraphics[width=\hsize]{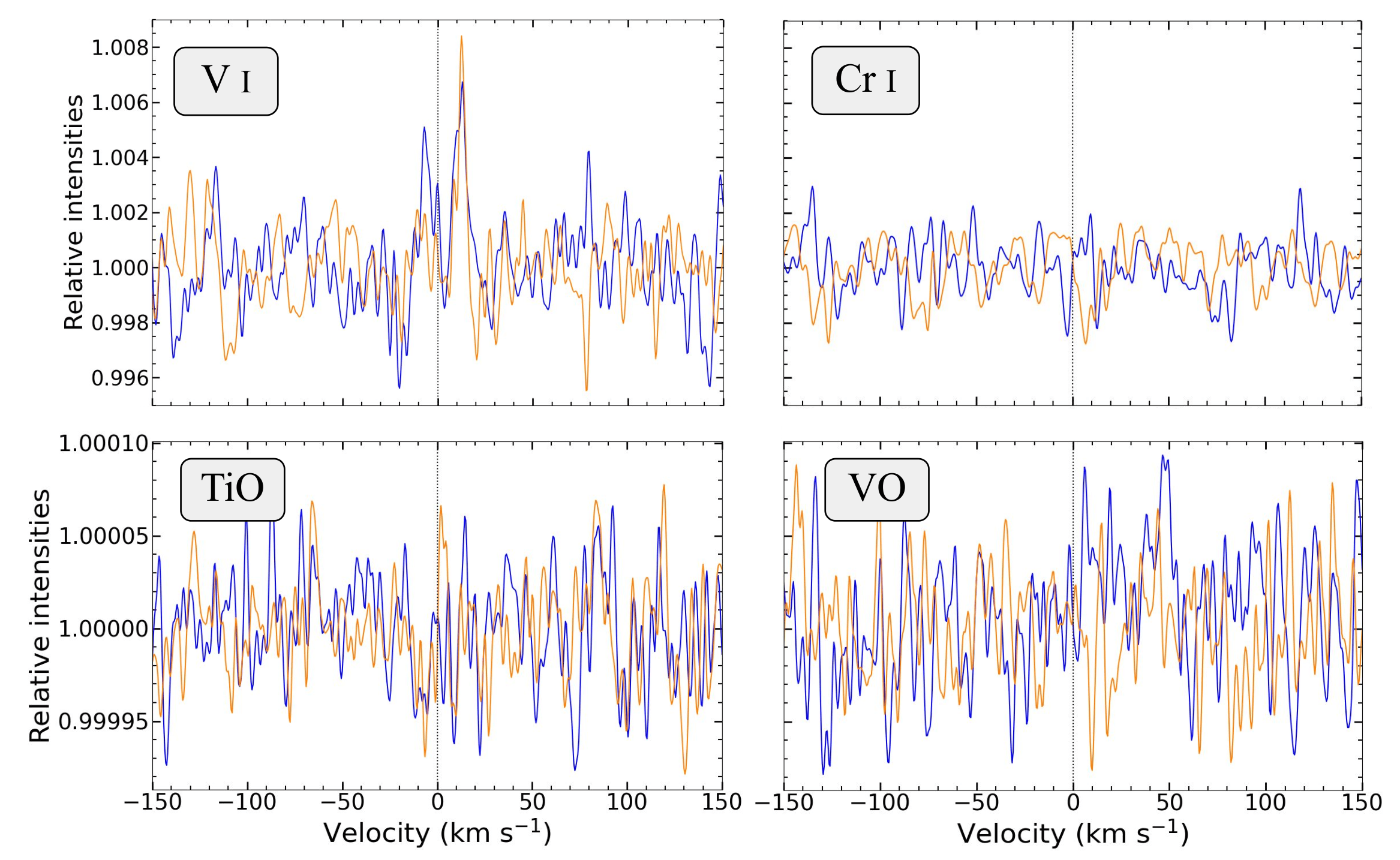}
    \caption{CCFs calculated at 1300 K for the combined in-transit planetary spectrum of non-detected species. Data from T1 and T2 are depicted in blue and orange, respectively. The dotted grey vertical line marks the planetary rest frame velocity. \textit{Top}: CCFs of the atomic species. \textit{Bottom}: CCFs of the molecular species.}
    \label{fig:CCF_non_detections}
    \end{figure}

    \begin{figure}[h!]
    \centering
    \includegraphics[width=\hsize]{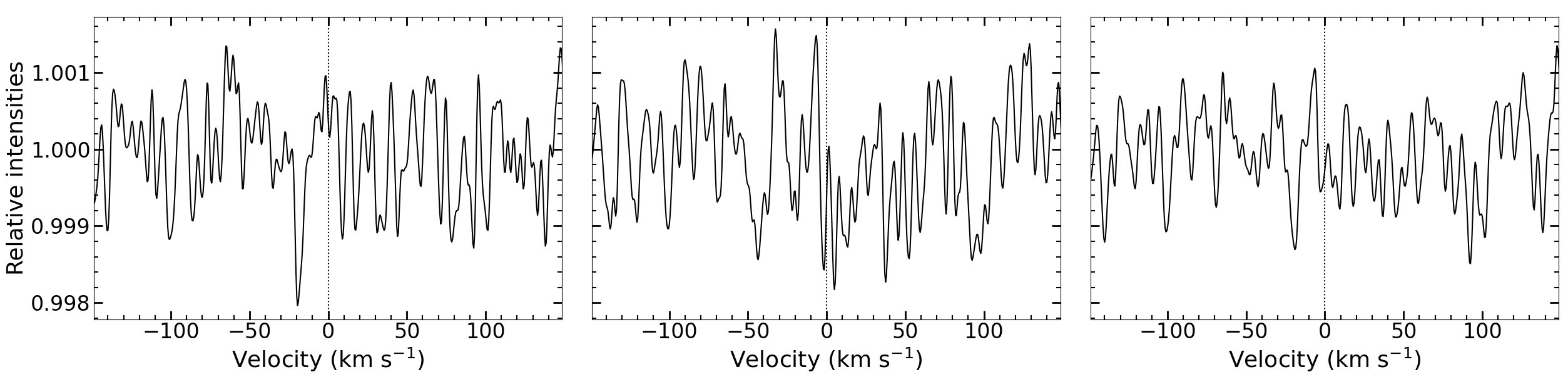}
    \caption{CCFs of the combined in-transit planetary spectrum calculated for Fe\,{\sc i} at a temperature of 1300 K. The dotted grey vertical line marks the planetary rest frame velocity. \textit{Left}: CCF of the first observing night. \textit{Middle}: CCF of the second night. \textit{Right}: CCF derived for the combined transmission spectrum of both nights.}
    \label{fig:Fe_1300_T1_T2_combined}
    \end{figure}

\pagebreak

\section{$K_{\mathrm{p}}$--$v_{\mathrm{sys}}$ maps}
\label{section:Kp-Vsys maps}

    \begin{figure}[h!]
    \centering
    \begin{minipage}{0.495\textwidth}
        \centering
        \includegraphics[width=\linewidth]{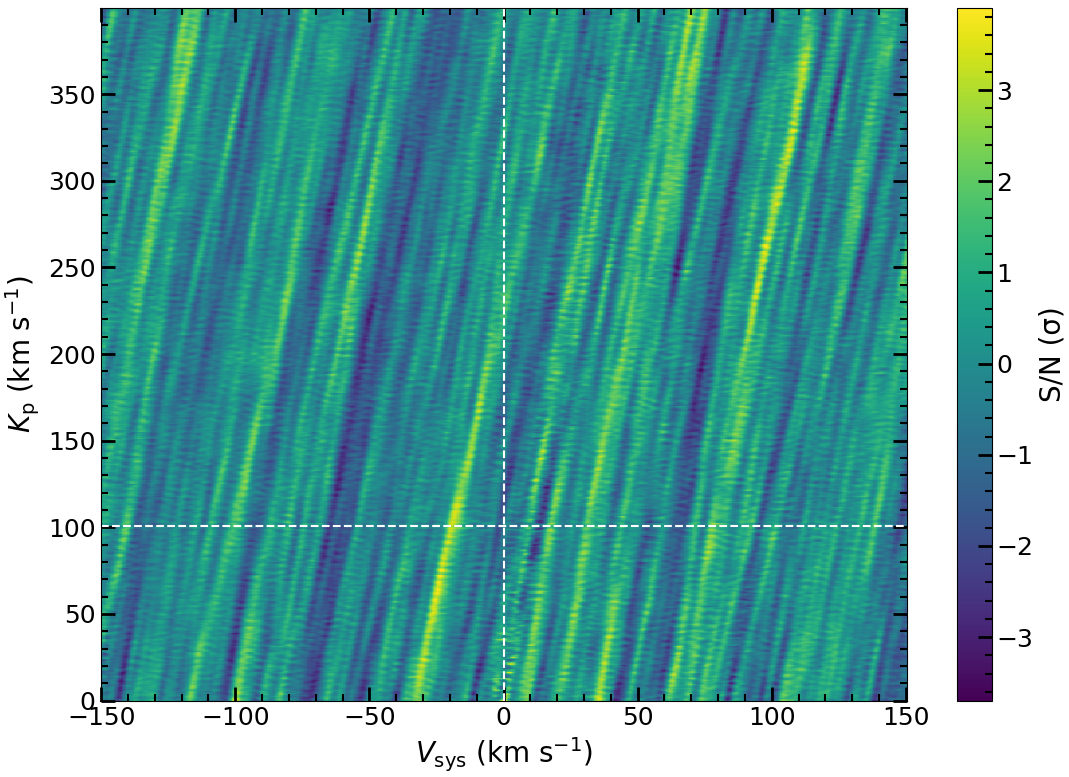}
        \caption*{}
    \end{minipage}
    \hfill
    \begin{minipage}{0.495\textwidth}
        \centering
        \includegraphics[width=\linewidth]{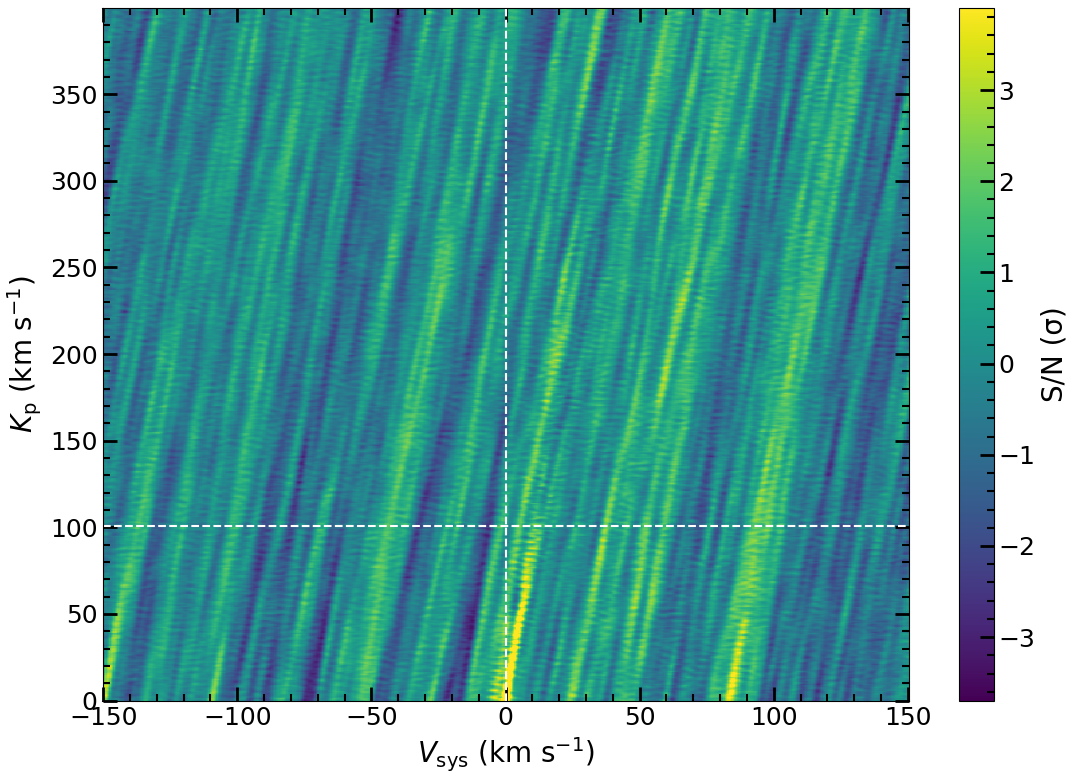}
        \caption*{}
    \end{minipage}
    \caption{$K_{\mathrm{p}}$–$v_{\mathrm{sys}}$ maps for the first transit (\textit{left}) and the second transit (\textit{right}). The horizontal white line marks the theoretical $K_{\mathrm{p}}$ value, and the vertical dashed line indicates the stellar rest-frame velocity. The CCFs were obtained using the Fe\,{\sc i} synthetic spectrum at 1300 K.}
    \label{fig:Kp_map}
    \end{figure}

\section{Injection-recovery tests for Na I}
\label{section:Injection-Recovery Tests for Na I}

    \begin{figure}[h!]
    \centering
    \includegraphics[width=0.6\hsize]{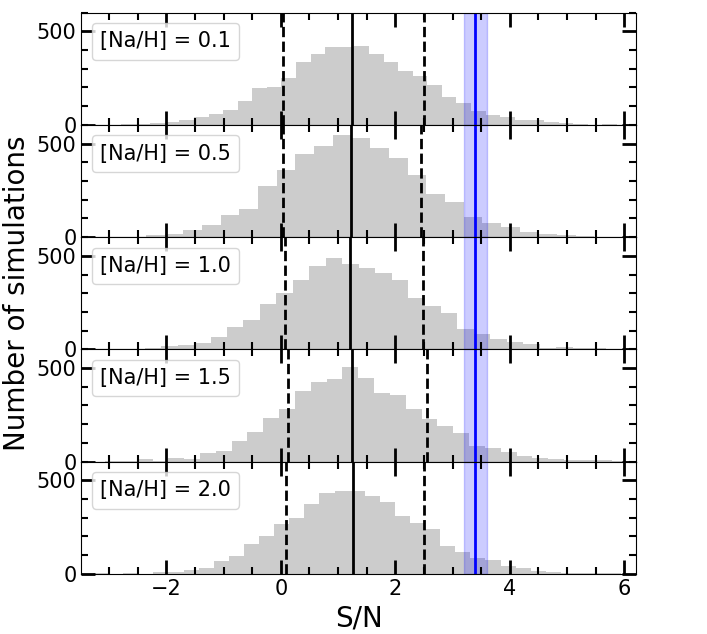}
    \caption{Results of the injection–recovery tests using Na\,{\sc i} synthetic spectra computed for different Na abundances. For easy comparison with Fig.~\ref{fig:histograms_Fe_syn_detection_T1}, the S/N of the tentative Fe\,{\sc i} detection in T1 is indicated by the blue band. For all the Na abundances explored, the recovered S/N remains below the S/N measured for the Fe\,{\sc i} signal in T1.}
    \label{fig:histograms_Na_syn_detection_T1}
    \end{figure}

\end{appendix}
\end{document}